\documentclass[letterpaper,dvipsnames]{article}
\pdfoutput=1
\usepackage{jheppub}
\usepackage{amsmath}
\usepackage{graphicx}

\usepackage{array}
\usepackage{ulem}
\usepackage{cancel}
\usepackage{xcolor}

\usepackage{slashed}
\usepackage{afterpage}
\usepackage{appendix}
\usepackage{multirow}
\usepackage{float}
\usepackage{mathtools}
\mathtoolsset{centercolon}

\usepackage{multirow}

\newcommand{\uv}{{\rm uv}}
\newcommand{\ir}{{\rm ir}}

\newcommand{\EM}{{\rm em}}

\newcommand{\sinw}{{\sin^2 (\theta_w)}}

\newcommand{\eff} {{ \rm eff}}

\newcommand{\aGUT}{\alpha_{\rm GUT}}
\newcommand{\GGUT}{\mathcal{G}}
\newcommand{\GUT}{{\mathbb{G}}}

\newcommand{\tr}{\text{tr}}

\newcommand{\ga}{g_{a \gamma \gamma }}
\newcommand{\ma}{m_a}
\newcommand{\gb}{g_{b \gamma \gamma }}
\newcommand{\mb}{m_b}

\title{Axion Couplings in Heterotic String Theory}

\author[a]{Prateek Agrawal,}
\author[b]{Michael Nee}
\author[a, c]{and Mario Reig}

\affiliation[a]{Rudolf Peierls Centre for Theoretical Physics, 
University of Oxford, Parks Road, Oxford OX1 3PU, United Kingdom}
\affiliation[b]{Department of Physics, Harvard University, Cambridge, MA, 02138, USA}
\affiliation[c]{Department of Physics, Royal Holloway University of London, Egham, Surrey, TW20 0EX, UK.}
\emailAdd{prateek.agrawal@physics.ox.ac.uk}
\emailAdd{mnee@fas.harvard.edu}
\emailAdd{mario.reiglopez@physics.ox.ac.uk}

\abstract{
We study the coupling of axions to gauge bosons in heterotic string theory. The axion-gauge boson couplings in the low energy 4d theory are derived by matching mixed anomalies between higher-form global symmetries and the zero-form gauge symmetry in the 10d theory. 
When the standard model gauge group is embedded in a single simple group in the 10d theory -- as is the case for almost all heterotic models studied in the literature --
the ratio of the axion-photon coupling to the axion mass is bounded above by the QCD line. This bound is relevant for a large number of axion searches which have sensitivity to axion parameter space above this line. The discovery of an axion in these searches will rule out a large class of heterotic models, making such a signal challenging to explain within heterotic string theory.}

\begin{document}

\maketitle

\section{Introduction}

\label{sec:introduction}

String theory is the best understood framework which unifies quantum mechanics and general relativity. The web of string dualities point to a unique string theory, leading to the hope that observable consequences may be derivable solely from the mathematical consistency of a fundamental theory. The vastness of the landscape of possible string vacua and only a partial understanding of string theory in various limits have made this dream impractical to achieve so far. In other words, the low-energy predictions of string theory depend on the details of the compactification, including the open questions of moduli stabilization, supersymmetry breaking and choice of local classical vacuum. In theories with light axions, however, the axion-photon coupling is quantized and is only sensitive to the topological data in the ultraviolet (UV) and unpolluted by intervening dynamics. 
In this work we show that, if measured, they can provide important constraints on the landscape of string theories.

Heterotic string theory~\cite{Gross:1984dd} provides a compelling framework which can reproduce the Standard Model (SM) at low-energies. Supersymmetric heterotic string theory is only consistent when the 10d gauge group is $SO(32)$ or $E_8\times E_8$, with the anomalies cancelled by the Green-Schwarz (GS) mechanism~\cite{Green:1984sg}\footnote{In non-supersymmetric scenarios, it is also possible to have consistent theories where the 10d gauge group is $SO(16) \times SO(16)$~\cite{Alvarez-Gaume:1986ghj}. It is also possible that there are emergent gauge symmetries when instantons shrink to zero size~\cite{Witten:1995gx, Benakli:1999yc}. We do not consider these cases in this work.}. There are many possible choices of compactification, the most prominent being Calabi-Yau (CY) manifolds~\cite{Candelas:1985en} and orbifolds~\cite{Dixon:1985jw, Dixon:1986jc, Ibanez:1986tp, Ibanez:1987xa}. Together with the large rank of the 10d gauge groups, this leaves a large number of possible symmetry breaking patterns~\cite{Witten:1985xc}. However, one of the reasons heterotic strings are so attractive for UV completing the standard model is that the standard model gauge group and number of chiral representations provide a direct and restrictive constraint on the possible compactification geometries. We will show in this work that the axion couplings provide a similar restriction on heterotic compactifications.

The presence of axions is seemingly a generic feature of string compactifications with sufficient complexity to describe the SM~\cite{Svrcek:2006yi, Arvanitaki:2009fg}. Axion masses are also protected by shift symmetries broken only by instanton effects, so they can be very light even when originating from physics at very high scales. This makes them potentially accessible to a wide range of astrophysical, cosmological and terrestrial experiments. This motivates the study of what restrictions on the axion parameter space may come from string theory, which is the focus of this paper.

Studying axions in string theory is further motivated from the bottom-up perspective by the quality problem. The question of how axions can be light is difficult to solve within a 4d effective theory where the shift-symmetry can be broken by local operators and coefficients of higher dimensional operators are generally considered to be $\mathcal{O}(1)$. Solutions to the quality problem within the 4d effective theory therefore must forbid operators up to very high dimension~\cite{Kamionkowski:1992mf} and often lead to tensions with early universe cosmology~\cite{Beyer:2022ywc,Lu:2023ayc}. In higher dimensions axions arise from higher-form fields integrated over compact cycles in the internal manifold. The shift symmetries which protect their masses are the 4d remnants of global higher form symmetries of the gauge theory in the higher dimensional space~\cite{Craig:2024dnl}. This breaking of the shift symmetry is exponentially suppressed by the action of a charged object wrapping the internal cycle which supports the axion. This leads to a shift symmetry of exponentially good quality, offering an elegant resolution to the quality problem.

Axion couplings to gauge bosons are uniquely sensitive to the details of high-energy physics which may be inaccessible to us through any other means. The reason for this is that they are quantised and therefore unaffected by renormalisation group flow, meaning they can be matched from the UV to the IR, offering a pristine window into the physics of the far UV~\cite{Agrawal:2017cmd,Fraser:2019ojt,Reece:2023iqn,Cordova:2023her,Agrawal:2023sbp,Choi:2023pdp}. As examples, the axion-photon coupling carries information about the minimal unit of electric charge~\cite{Agrawal:2019lkr,Yin:2023vit}, as well as possible unification of the SM in the UV~\cite{Agrawal:2022lsp}. In heterotic models, the axion couplings to gauge bosons in 4d come from the mixed anomalies between higher-form global and gauge symmetries in the 10d theory. To derive the axion couplings in the effective theory we identify the relevant higher-form symmetries in the 10d theory and match these anomalies from the 10d supergravity theory to the 4d EFT.

The axion-photon coupling and mass are the two main parameters relevant for the experimental search for axions. The main result of this work is that in many heterotic models there is a bound relating these two quantities:
\begin{equation}\label{eq:bound_coupling}
    \frac{g_{a\gamma  \gamma}}{m_a}\leq C \, \frac{\alpha_{\EM}}{2\pi}\frac{1}{m_\pi f_\pi}\, ,
\end{equation}
which is satisfied by every axion $a$ in the theory. In equation~\eqref{eq:bound_coupling} $C$ is a calculable $\mathcal{O}(1)$ coefficient whose value depends on the model under consideration. Axions saturating this bound are said to lie on the QCD line. This result comes from the fact that in a theory where the SM comes from a simple gauge group in the UV, any axion couples to photons and to gluons with comparable strength, which generates an axion mass after QCD confines~\cite{Agrawal:2022lsp}. This result is independent of both the details of gauge symmetry breaking and the intervening physics between the IR and UV scales, meaning that experiments searching for light axions through the photon coupling are sensitive to the UV completion of the SM.

The key feature of heterotic string theories that determines the low energy axion-photon coupling is how the SM gauge group is embedded into the 10d gauge theory. We show that for the $SO(32)$ theory there is always a bound of the form~\eqref{eq:bound_coupling}, while for the $E_8 \times E_8$ theory an axion evading~\eqref{eq:bound_coupling} can only be found in theories where either QCD or hypercharge is a diagonal subgroup of factors coming from both $E_8$'s. The bound applies to all models where the SM gauge group is embedded into one $E_8$ factor, including all models considered in references~\cite{Ibanez:1987sn, Font:1989aj,Choi:2006qj,Choi:2009jt,Blumenhagen:2005ga,Anderson:2011ns}.

The vast majority of axion searches rely on detecting axions through their photon coupling~\cite{Sikivie:1983ip,Sikivie:1985yu,Wilczek:1987mv,CAST:2004gzq, Kahn:2016aff, Arvanitaki:2017nhi, Baryakhtar:2018doz, Chaudhuri:2018rqn, Fedderke:2019ajk,Langhoff:2022bij,QSHS:2023jny}, although alternative detection channels are possible~\cite{Moody:1984ba, Graham:2011qk, Budker:2013hfa, Graham:2013gfa, Armengaud:2013rta, Arvanitaki:2014dfa, Arvanitaki:2021wjk, Berlin:2022mia}.  The axion mass determines which type of experiments will be sensitive to a signal (see~\cite{Graham:2015ouw,Irastorza:2018dyq,OHare:2024nmr} for reviews). The restrictions we find on the axion-photon coupling are relevant to axion searches sensitive to parameter space above the QCD line. These range from cosmological probes~\cite{Minami:2020odp,BICEPKeck:2021sbt,SPT-3G:2022ods,Eskilt:2022cff,Nakai:2023zdr,Diego-Palazuelos:2023mpy,Cosmoglobe:2023pgf, POLARBEAR:2023ric, POLARBEAR:2024vel} and superradiance~\cite{Arvanitaki:2010sy}, $\gamma$-ray telescopes~\cite{Wouters:2013hua, Marsh:2017yvc,Reynolds:2019uqt, Reynes:2021bpe}, haloscopes and helioscopes~\cite{Marsh:2018dlj, Lawson:2019brd, Beurthey:2020yuq,Schutte-Engel:2021bqm,DMRadio:2022pkf,Aja:2022csb,Bourhill:2022alm,ALPHA:2022rxj,ADMX:2023rsk,Oshima:2023csb,DeMiguel:2023nmz,Ahyoune:2023gfw,Alesini:2023qed,BREAD:2023xhc,CAST:2024eil,Friel:2024shg, Kalia:2024eml}\footnote{There are also cosmological~\cite{Cadamuro:2011fd,Depta:2020wmr} and collider~\cite{Bauer:2017ris} searches sensitive to heavier axion masses to the right of the QCD line.}. Moreover, the recently reported (tentative) detection of a birefringence signal may already be an indication of a very light axion coupled to photons~\cite{Minami:2020odp,Eskilt:2022cff,Nakai:2023zdr,Diego-Palazuelos:2023mpy,Cosmoglobe:2023pgf}. This means if any signal were observed in these experiments, it would have significant implications for heterotic model building. It would tell us that if the UV completion of the SM was heterotic string theory, then the UV gauge group is $E_8 \times E_8$, and the SM hypercharge is contained within both $E_8$ factors.
\\

The ideas presented in this paper have some overlap with the swampland program, which posits that not all EFTs that are consistent from the low-energy point of view remain so when coupled to gravity~\cite{Vafa:2005ui}. The many swampland conjectures restrict the range of possible IR theories by determining general properties of the quantum gravity landscape, and are based on black hole arguments or examples from well-understood corners of string theory~\cite{Arkani-Hamed:2006emk, Ooguri:2006in, Banks:2010zn, Ooguri:2016pdq}. In this work we use anomaly matching to derive robust constraints on possible axion couplings within a large class of UV completions. The restrictions we find on these couplings then offers a way to experimentally probe large swathes of the string landscape itself.

Axions in heterotic string theory have been well-studied in the past~\cite{Witten:1984dg,Kim:1988dd,Choi:1985bz,Choi:1985je,Derendinger:1985cv,Banks:1996ss,Choi:1997an,Choi:2011xt,Buchbinder:2014qca,Choi:2014uaa,Im:2019cnl}, with a particular emphasis given to the overabundance problem associated with large axion decay constants~\cite{Svrcek:2006yi}.
Instead of focusing on the magnitude of the axion decay constant, which we remark is a dynamical aspect of the theory, in this paper we study in detail the axion-photon coupling and its topological nature. Our main results are independent of the value of the axion decay constant, although large decay constants make axions much more difficult to observe in experiments. There is also a large body of literature on axions in more general string contexts, in particular type IIB string theory~\cite{Arvanitaki:2009fg,Conlon:2006tq,Cicoli:2012sz,Hebecker:2018yxs,Demirtas:2018akl,Halverson:2019cmy,Mehta:2021pwf,Demirtas:2021gsq,Foster:2022ajl,Gendler:2023hwg, Gendler:2023kjt, Reece:2024wrn,Gendler:2024adn}.
\\

This paper is organised as follows: in section \ref{sec:4d_GUTs} we review the constraints that unifying the SM in 4 dimensions imposes on axion couplings. Section \ref{sec:gaugegroups} is dedicated to reviewing the mechanisms of symmetry breaking in heterotic string theory, and how the effective 4d gauge symmetry is obtained from the 10D gauge group. Sections \ref{sec:anomalymatching}, \ref{sec:heterotic_axions} and \ref{sec:4d_eft_pheno}  contain the main results. In section \ref{sec:anomalymatching} we show how to get the axion couplings in 4d by matching the mixed anomalies of the higher-form symmetries in 10d to the mixed anomalies of the axion shift-symmetries in the 4d EFT. In section~\ref{sec:heterotic_axions} we derive the effective theory describing axion physics in 4d. In section~\ref{sec:4d_eft_pheno} we describe the phenomenology of axions in heterotic string theory and the implications for axion experiments are summarised in section \ref{subsec:experiments}.

\section{Review of Axions in 4d GUTs}
\label{sec:4d_GUTs}
If the SM descends from a 4d theory with a simple unified group $\GUT$, the axion couplings to photons were shown to be very restrictive in~\cite{Agrawal:2022lsp}. The axion must be a singlet under the gauge symmetry, and couples to the full UV gauge group as:
\begin{align}
    \mathcal{L} = \mathcal{A} \frac{\aGUT}{2\pi } \frac{a}{F_a}  \tr ( \GGUT_{\mu\nu} \tilde \GGUT^{\mu\nu})\,,
\end{align}
where $\GGUT$ is the unified field strength, $\aGUT$ the GUT coupling, and $2\pi F_a$ the periodicity of the axion field. In theories with multiple axions, we define $a$ as the linear combination that saturates the mixed anomaly between the global $U(1)$ symmetry of the axion with the GUT gauge group.

The coefficient $\mathcal{A}$ is an integer anomaly coefficient and can be used to match axion couplings in the UV and in the IR. In the IR, the axion couplings to photons and gluons are parameterised by the rational numbers $E$ and $N$,
\begin{align}\label{eq:4d_axion_couplings}
    \mathcal{L} 
    \supset 
    \frac{a}{4\pi F_a} 
    \left[ \alpha_\EM E F_{\mu\nu} \tilde F^{\mu\nu}
    + \alpha_s N G^a_{\mu\nu} \tilde G^{a, \mu\nu})
    \right] \, .
\end{align}
Once the embedding of the SM into the unified gauge group is specified, the ratio $E/N$ is fixed\footnote{For most viable 4d unified theories the ratio is $E/N=8/3$. See the appendix of \cite{Agrawal:2022lsp} for more general cases.}. The ratio $E/N$ is related to the global structure of the SM gauge group~\cite{Reece:2023iqn,Cordova:2023her,Agrawal:2023sbp,Choi:2023pdp}.

Below the QCD scale, the axion gets a potential and the axion-photon coupling gets corrections from mixing with the pion, 
\begin{align}\label{eq:4d_axion_couplings2}
    \mathcal{L} 
    \supset 
    -\frac14 g_{a\gamma\gamma}
     a F_{\mu\nu} \tilde F^{\mu\nu}
     - V(a)\,,
\end{align}
where
\begin{align}
    \ga
    &= \frac{\alpha_\EM  }{\pi F_a} \left( E -1.92 N \right), 
    \qquad
    V_{\rm QCD}(a)=
    -f_\pi^2 m_\pi^2
    \sqrt{1-\frac{4z}{(1+z)^2} \sin^2\left(N\frac{a}{F_a}\right)},
    \qquad z = \frac{m_u}{m_d} \, .
    \label{eq:QCDcoupling}
\end{align}
These QCD axion parameters saturate the bound~\eqref{eq:bound_coupling}, with the undetermined coefficient $c$ being given by
\begin{align}
    C = \left( \frac{E}{N} -1.92 \right) \left(\sqrt{z} + \frac{1}{\sqrt{z}}\right)\,.
    \label{eq:const_qcd}
\end{align}

Axions that do not shift under the unique anomalous $U(1)$ symmetry do not have direct couplings to gauge bosons, but can inherit a coupling through kinetic or mass mixing effects parameterised by a mass mixing matrix. The solution of the strong CP problem requires that the mass matrix of axions have a zero eigenvalue, so that the corresponding eigenvector only obtains mass from QCD and is identified as the QCD axion. The orthogonal linear combinations inherit their coupling from the anomalous axion, so the ratio of their couplings to photon and gluons is still fixed by the ratio $E/N$. If their mass eigenvalues $m_b$ are smaller than that of the QCD axion, the photon couplings are suppressed relative to those of the QCD axion by the mixing parameter $m_{b}^2/\ma^2$. For $m_b > \ma$, the couplings remain unsuppressed but the axions are heavier than the QCD axion. In either case, these additional axion have photon coupling $\gb$ and mass $\mb$ which satisfy equation~\eqref{eq:bound_coupling}, falling to the right of the QCD line in parameter space plots.

The strict result~\eqref{eq:bound_coupling} holds for theories where the SM is unified into a simple gauge group in the UV. In theories with only partial unification such as flipped $SU(5)$~\cite{Barr:1981qv,Derendinger:1983aj}, Pati-Salam~\cite{Pati:1974yy} and trinification~\cite{Babu:1985gi} models, these predictions may not hold.  In that case, unless extra symmetries are imposed the GUT predictions of the SM quantum number and gauge coupling unification are lost. Another class of generalizations of GUTs are Orbifold GUTs \cite{Kawamura:1999nj,Hall:2001pg, Hall:2001tn, Hall:2001xb, Hebecker:2001jb, Hebecker:2001wq,Altarelli:2001qj}, arising from higher-dimensional constructions of unified theories with symmetry breaking using the topology or boundary conditions in the extra-dimensions. The attractive feature of these constructions is that they allow for some predictions of grand unification to remain, such as gauge coupling unification, while ameliorating some of the unfavourable features of simple GUTs, like the doublet-triplet splitting problem. In general, these theories allow for incomplete GUT multiplets, so the argument given above for axion-photon couplings in simple GUT gauge groups does not apply straightforwardly.

Remarkably, all these cases are present in geometric compactifications of heterotic string theory~\cite{Witten:1985xc}. Most compactifications that have been studied reproduce the axion phenomenology of 4D simple GUTs, but include the desirable features of orbifold GUTs! It is in principle also possible to also produce the partially unified groups such as flipped SU(5), but it is hard to construct a realistic model of this sort. In the next sections we discuss these possibilities in detail.

\subsection{Embedding the Standard Model}
\label{subsec:embedding}
One main input that governs the low-energy axion couplings is how the SM is embedded in $\GUT$. A non-Abelian subgroup $\mathbb{H}\subset \GUT$ has embedding level $k \in \mathbb{Z}_{>0}$  defined by
\begin{align}
 k 
 &=
  \frac{\sum_i \mu_{R_i}}{\mu_R}\,,
\end{align}
where the representation $R$ of $\GUT$ decomposes into representations $R_i$ of $\mathbb{H}$, and $\mu_R$ is the Dynkin index of the representation, $\mu_R \delta^{ab}= \tr_R t^a t^b$. The level of embedding is independent of the choice of representation in the definition above. A simple example is obtained by taking $\mathbb{H}$ to be the diagonal subgroup of a product group $\mathbb{H}_1\times \ldots \times \mathbb{H}_k$ contained in $\GUT$:
\begin{align}
    \GUT \supset \mathbb{H}_1\times \ldots \times \mathbb{H}_k \supset \mathbb{H}= \mathbb{H}_{\rm diag} \, .
\end{align}
In this case, the matching condition for the coupling, $\alpha_\ir$, of the low energy gauge group is
\begin{align}
    \frac{1}{\alpha_\mathbb{H}} = \sum_{i=1}^k \frac{1}{\alpha_{\mathbb{H}_i}} \, ,
    \label{eq:matching}
\end{align}
so for unified groups in the UV, the IR coupling $\alpha_\mathbb{H}$ is smaller than $\alpha_\GUT$ by the factor $k$. For Abelian subgroups the embedding level depends on the normalisation of the $U(1)$ generator. It is a positive rational number and the coupling satisfies the matching condition $\alpha_\ir^{-1} = k_1 \alpha_\uv^{-1}$, where $\alpha_\ir \, (\alpha_\uv)$ are the couplings below (above) the matching scale. The embedding of the SM is therefore described by the levels $(k_3, k_2; k_1)$, where $k_{2},\, k_3 \in \mathbb{Z}_{>0}$ and $k_{1} \in \mathbb{Q}_{>0}$. 

For a unified theory both the value of $\sin^2 (\theta_w)$ at the GUT scale and the ratio of anomaly coefficients $E/N$ in the axion coupling are determined solely by the embedding levels:
\begin{align}
    \label{eq:sinw}
    &\sinw = \frac{k_2}{k_1 + k_2} \, ,  
    \qquad 
    \frac{E}{N} = \frac{k_1 + k_2}{k_3} \, .
\end{align}
For 4d GUT theories the standard choice of embedding is $(k_3, k_2; k_1) = (1,1; 5/3)$. This choice correctly predicts the value of the weak mixing angle, explains why colour singlets have integer electric charges, and leads to $E/N= 8/3$. In heterotic compactifications the standard choice of embedding leads either to unbroken $SU(5)$ in 4d (with no representations to break it to the SM), or the presence of fractional charges from twisted states through Wilson line breaking \cite{Schellekens:1989qb}, as discussed in section~\ref{subsec:het-embedding}. 

\section{Gauge groups in heterotic string theory}
\label{sec:gaugegroups}
Heterotic and type-I string models have been extensively studied as potential UV completions of the standard model. In the weakly coupled geometrical regime and assuming supersymmetry, they are described by 10d supergravity coupled to a non-Abelian gauge theory with only two possible gauge groups: $E_8 \times E_8$ or $SO(32)$. If we give up on spacetime supersymmetry, it also possible for the UV gauge group to be $SO(16) \times SO(16)$~\cite{Alvarez-Gaume:1986ghj}, however we do not consider that case here. These gauge groups are suitable starting points for building realistic low-energy models of particle physics, but must be broken down to the standard model by gauge field backgrounds in the internal dimensions. In this section we review the different symmetry breaking mechanisms available in string models and their implications for the low energy gauge group and massless chiral spectrum. Our discussion will be mostly general, however, where relevant we specialise to the case of a $E_8 \times E_8$ gauge group.

For the supersymmetric models we consider, the relevant part of the bosonic supergravity action in 10d reads:
\begin{equation}\label{eq:sugra_10}
    S_{10}=\frac{1}{2\kappa_{10}^2}\int e^{-2\phi} \left [ d^{10}x \sqrt{-g_{10}} R +4 d\phi \wedge \star d\phi
    -\frac{l_s^4}{2} H \wedge \star H -\frac{\alpha^\prime}{4}\tr (\mathcal{F} \wedge \star \mathcal{F})\right] - \frac{1}{768\pi^3}{ \int B\wedge X_8}\,,
\end{equation}
where $\phi$ is the dilaton, $R$ is the Ricci scalar, $\mathcal{F}$ is the gauge field strength, $\tr$ represents the trace in the fundamental representation and we adopt the conventions of ref.~\cite{Svrcek:2006yi} in normalising the fields. In addition to the Yang-Mills sector and gravity, $S_{10}$ includes the gauge invariant field strength of the two-form field, $B$:
\begin{equation}
    H= dB + \frac{1}{16\pi^2}\left (\omega^{\rm grav}_3 - \omega_3\right )\,.
\end{equation}
$\omega^{\rm grav}_3$ and $\omega_3$ are the Chern-Simons forms for the gravity and gauge sectors: 
\begin{align}
    &\omega^{\rm grav}_3 = \tr \left( \Omega \wedge d \Omega + \frac{2}{3} \Omega \wedge \Omega \wedge \Omega \right)  \, ,
    &&\omega_3 = \tr \left( A \wedge dA - \frac{2i}{3} A \wedge A \wedge A \right) \, ,
\end{align}
where $\Omega$ is the spin connection. They satisfy $d\omega_3  = \tr \mathcal{F} \wedge \mathcal{F} , \, d\omega^{\rm grav}_3  = \tr \mathcal{R} \wedge \mathcal{R}$, where $\mathcal{R}$ is the Riemann curvature two-form. The Bianchi identity for the gauge invariant field strength is therefore:
\begin{equation}
    dH = \frac{1}{16\pi^2} \left( \tr\mathcal{R}^2 - \tr \mathcal{F}^2 \right)\, ,
    \label{eq:bianchi}
\end{equation}
where we use the shorthand $\tr \mathcal{R}^2 = \tr \mathcal{R} \wedge \mathcal{R}$. For the $E_8 \times E_8$ model $\tr \mathcal{F}^2 = \tr_1 \mathcal{F}^2 + \tr_2 \mathcal{F}^2 $ where the subscripts indicate traces over each separate $E_8$ factor. This identity states that $\tr \mathcal{R}^2$ and $\tr \mathcal{F}^2$ differ by at most an exact form, relating the background $\mathcal{F}$ and $\mathcal{R}$. The last term in Eq.(\ref{eq:sugra_10}) ensures that anomalies cancel~\cite{Green:1984sg} and provides axion couplings in the 4d EFT. The precise form of $X_8$ depends on the gauge group in 10d, which we discuss further in section~\ref{sec:anomalymatching}. We consider cases where the 10d manifold, $\mathcal{M}_{10}$, is a product of 4d Minkowski space $M_4$ with a compact internal space $K$:
\begin{align}
    \mathcal{M}_{10} = M_4 \times K \, .
\end{align}

\subsection{Breaking the 10D gauge group}
\label{sec:reducing_10D_group}
Neither $E_8$ nor $SO(32)$ contain chiral representations, so these groups must be broken to a smaller subgroup in a topologically non-trivial way if we are to recover the SM at low energies. In this section we briefly review the different ways the 10d gauge group $\mathbb{G}$ may be broken to a subgroup at the compactification scale -- these being breaking by background gauge fields, orbifold projections and Wilson lines. The first two mechanisms can lead to a chiral fermion spectrum in the EFT below the compactification scale, while Wilson line breaking does not change the massless fermion spectrum. A brief summary of each of the mechanisms is presented here, while some specific examples are given in the later sections.

\subsubsection{Non-trivial internal gauge backgrounds}
\label{subsec:explicit}
The first breaking mechanism we consider is turning on background gauge fields, $\bar{\mathcal{F}}$, in the internal space~$K$. The gauge group of the effective theory below the compactification scale will then be the subgroup which commutes with all of the background fields. We can expand $\bar{\mathcal{F}}$ as
\begin{align}
    \bar{\mathcal{F}} = \sum_a \bar{\mathcal{F}}^a \bar{t}^a \, ,
\end{align}
where we define the sum on the right hand side to be over only the generators $\bar t^a$ of $\mathbb{G}$ for which $\bar{\mathcal{F}}^a \neq 0$. Then the generators $t^a$ which parameterise the gauge group of the effective theory are those which satisfy:
\begin{align}
    \left [ \bar{t}^a, t^b \right ] = 0  \qquad \forall \, a \, .
    \label{eq:commutant}
\end{align}
The background fields must satisfy the Bianchi identity~\eqref{eq:bianchi}, so the possibilities for $\bar{\mathcal{F}}$ depend on the choice of internal manifold $K$. This is as the background $\bar{\mathcal{R}}$ is the field strength associated with the holonomy group of $K$, so manifolds $K$ with different holonomy groups will allow for different choices of $\bar{\mathcal{F}}$.

\subsubsection{Orbifolds}\label{sec:het_orbifolds}
An alternative way of breaking the 10d gauge group is by orbifolding \cite{Dixon:1985jw, Dixon:1986jc, Ibanez:1986tp, Ibanez:1987xa}. In this case the starting point is an internal manifold $K_0$ that is flat, typically a torus (the Bianchi identity is then satisfied for $\tr \mathcal{F}^2 =0$). Then we consider the quotient $K_0/\mathbb{F}$, where $\mathbb{F}$ is a discrete symmetry group of $K_0$. It is important that $\mathbb{F}$ does not act freely on $K_0$, i.e. there are fixed points $x_0 \in K_0$ such that $\mathbb{F}:x_0 \to x_0$. The fields on $K_0$ must also be given some charges under the action of $\mathbb{F}$; in particular, the gauge fields will transform as 
\begin{align}
   A (x) \to P_f A ( f(x)) \, ,
   \label{eq:orb_action}
\end{align}
where the $P_f$ are matrices in the adjoint representation of $\mathbb{G}$ which form a representation of $\mathbb{F}$. For example, if $\mathbb{F} = \mathbb{Z}_N$, then the $P_f$ must satisfy $ (P_f)^N = 1$. The gauge fields which have a non-trivial transformation under the orbifold action~\eqref{eq:orb_action} are removed from the massless spectrum, leaving a reduced 4d gauge group consisting of the fields with trivial transformation law under~\eqref{eq:orb_action}. In some cases orbifold theories are related to compactifications on smooth manifolds by a `blow-up' procedure involving replacing the singular points with manifolds of the correct holonomy, although this is not the most general possibility~\cite{Narain:1986qm}.

\subsubsection{Wilson Line Breaking}
\label{subsec:Wilson}
Both the above methods of breaking the gauge group have the important feature that the fermions in the 4d theory come in representations of the broken gauge group. In particular, complex representations and a chiral fermion spectrum become possible. However, if the internal manifold $K$ is not simply connected, $\pi_1(K) \neq 0$, then the gauge group can be further broken by expectation values of Wilson lines~\cite{Candelas:1985en,Witten:1985xc}
\begin{equation}
    W^a=\exp \left( i \int_\gamma A^a \right)\,,
    \label{eq:Wilson}
\end{equation}
which wrap internal cycles $\gamma$ in $K$. The difference between the gauge field backgrounds discussed in the previous section is that the field strength $\mathcal{F}$ vanishes, so the background for $A^a$ can be gauged away locally. Wilson line backgrounds break the gauge group to the subgroup which commutes with the $W^a$, i.e. the subgroup satisfying a relation analogous to~\eqref{eq:commutant}. Wilson line breaking does not change the number of chiral fermions in the low energy theory. 

\subsection{Breaking to the SM}
\label{subsec:het-embedding}
Following the discussion of section~\ref{subsec:embedding}, the SM gauge groups will be embedded in the 10d gauge group(s) with some embedding levels $k_1, \, k_2, \, k_3$. However, there are some interesting implications of the breaking mechanisms in string models which mean they behave differently to higgsed 4d theories; we review these differences in this section. 

\subsubsection{Embedding in a simple group}
\label{subsec:GUTembedding}
In this section we consider the case where the SM is embedded into a simple gauge group in the UV, leaving the more general case for section~\ref{subsec:general_embedding}. This therefore applies when the SM comes from a single $E_8$ factor or if the UV gauge group is $SO(32)$. The scenario we have in mind is where the 10d gauge group $\mathbb{G}$ is broken to a unified gauge group $\mathbb{G}_{\rm GUT}$ either by background gauge fields or orbifolding, then further down to the SM by Wilson lines. 

A significant difference between 4d GUTs and their stringy counterparts is the absence of a GUT breaking mechanism in 4d if the SM is embedded in the usual way (with $(k_3, k_2; k_1) = (1,1,5/3)$ and integer electric charges). This is due to the fact that if all colour singlets have integer charges with this embedding, then the low energy gauge group must contain an unbroken $SU(5)$~\cite{Schellekens:1989qb}. This implies that there must either exist fractional charges or the embedding of the SM is non-standard. For the standard embedding, breaking $SU(5)$ to the SM can be achieved by Wilson lines along the internal dimensions. These necessarily introduce twisted states with fractional charges \cite{Wen:1985qj}, although these are generically heavy so typically don't pose a problem for phenomenology. 

If the SM is embedded in a non-standard manner then $\sinw \neq 3/8$ (see equation~\eqref{eq:sinw}). This would require the presence of additional matter to influence the running of the SM gauge couplings so that they reach their measured values in the IR. The allowed values of $\sin^2\theta_w$ are not completely arbitrary, however, but are instead restricted by requiring we reproduce the SM fermion content. Having three generations requires that $k_2$ and $k_3$ must either be 1 or 3 and the existence of the lepton singlet requires $k_1 \geq 1$~\cite{Schellekens:1989qb, Font:1990uw}. 

In this class of models, the usual GUT relations between Yukawa coupling constants do not necessarily apply, offering interesting solutions to old problems of unified theories~\cite{Witten:1985xc}. Wilson line breaking does not change the chiral index, so the matter fields fill out full representations of $\mathbb{G}_{\rm GUT}$ even though they do not arise from a single $\mathbb{G}_{\rm GUT}$ multiplet in the UV theory. For example, for an $SU(5)$ theory with two $\bar 5 \sim (\bar d, L)^T$ multiplets, the $\bar d$ component may be made heavy in one while the $L$ is made heavy in the other. The light fermions then fill out a full $\bar 5$ but the Yukawa couplings of the $\bar d$ and the $L$ are unrelated as they came from different UV multiplets.

\subsubsection{General embeddings}
\label{subsec:general_embedding}
The more general possibility is where the UV gauge group is $E_8 \times E_8$ and the SM is embedded non-trivially into both $E_8$ factors. Similar to the 4d GUT case reviewed in section~\ref{sec:4d_GUTs}, this embedding will be a key input to axion couplings to gauge bosons in the effective theory. In preparation for later sections, here we relate the $E_8 \times E_8$ gauge fields in the 10d theory to the SM gauge fields and establish notation which we will use throughout the later sections. 

In general each of the SM gauge groups will have two embedding levels $k^{(i)}$, where $i= 1, \, 2$ refers to the $E_8$ factors. Where relevant we will use $k^{(i)}_1, \, k^{(i)}_2,$ \& $ k^{(i)}_3$ to denote the embedding levels of $U(1)_Y, \, SU(2)_L,$ and $SU(3)_c$ in the $i-$th $E_8$ factor, respectively, and $ k^{(i)}_j$ for $j \geq 4$ to refer to any additional gauge groups which may be present in the EFT at the compactification scale. At this scale the 4-forms in 10 dimensions descend to four forms of the 4d gauge field strengths $F^{(i)}_j$:
\begin{align}
    &\tr_1 \left(\mathcal{F} \wedge  \mathcal{F} \right) 
    \to \sum_j k^{(1)}_j \tr  \left(F^{(1)}_j \wedge  F^{(1)}_j  \right) \, ,
    &&\tr_2 \left(\mathcal{F} \wedge  \mathcal{F} \right) 
    \to \sum_j k^{(2)}_j \tr \left( F^{(2)}_j \wedge  F^{(2)}_j \right) \, ,
    \label{eq:embeddinge8}
\end{align}
where the sum is over all the gauge fields which survive the breaking mechanisms discussed earlier in this section. 

We can now ask how to relate the $E_8$ traces to the SM gauge groups. If the SM contains contributions from both $E_8$ factors then the orthogonal linear combinations of gauge fields must be broken. If hypercharge, for example, is embedded into both $E_8$ factors then we can choose our basis so that $F_Y$ appears as a linear combination of $F^{(1)}_1$ and $F^{(2)}_1$:
\begin{align}
    &F_Y = \cos (\theta_1) F^{(1)}_1 + \sin (\theta_1)  F^{(2)}_1 \, ,
    && \tan(\theta_1) = \sqrt{k^{(2)}_1 / k^{(1)}_1} \, .
\end{align}
Denoting the orthogonal linear combination by $\bar F$ we can invert this relation to get:
\begin{align}
    &F_1 = \cos(\theta_1) F_Y - \sin (\theta_1) \bar F \, ,
    &&F_2 = \sin(\theta_2) F_Y + \cos (\theta_2) \bar F \, .
\end{align}
Performing a similar procedure for the QCD and weak gauge fields (with field strengths $G$ and $W$) we can write the $E_8$ traces as: 
\begin{equation}
\begin{aligned}
    \tr_1 \left(\mathcal{F} \wedge \mathcal{F} \right) 
    &= \frac{(k^{(1)}_1)^2}{k^{(2)}_1 + k^{(1)}_1} F_Y \wedge  F_Y
    +\frac{(k^{(1)}_2)^2}{k^{(2)}_2 + k^{(1)}_2} \tr \left( W\wedge W\right)
    +\frac{(k^{(1)}_3)^2}{k^{(2)}_3 + k^{(1)}_3} \tr \left( G\wedge G\right)
    + \ldots    \, ,
    \\
    \tr_2 \left(\mathcal{F} \wedge \mathcal{F} \right) 
    &= \frac{(k^{(2)}_1)^2}{k^{(2)}_1 + k^{(1)}_1} F_Y \wedge  F_Y
    +\frac{(k^{(2)}_2)^2}{k^{(2)}_2 + k^{(1)}_2} \tr \left( W\wedge W\right)
    +\frac{(k^{(2)}_3)^2}{k^{(2)}_3 + k^{(1)}_3} \tr \left( G\wedge G\right)
    + \ldots \, ,
\end{aligned}
\label{eq:embed10dto4d}
\end{equation}
where the `$\ldots$' refer to possible additional gauge groups appearing on the right hand side. Irrespective of the mechanism which breaks the 10d gauge theory to 4d, equation~\eqref{eq:embed10dto4d} describes the most general way the SM can be embedded into $E_8 \times E_8$.

\section{Matching Anomalies from 10d to 4d}
\label{sec:anomalymatching}
In this section we derive the anomaly coefficients which describe the mixed anomalies between the axion shift symmetries and the 10d gauge group. These coefficients, combined with the embedding levels $k^{(i)}_j$ defined in the previous section, are ultimately what determine the axion couplings to the SM. The key result that we emphasize is that the low-energy axion couplings are topological, and hence robust to dynamical effects such as RG flow. In four dimensions this statement is tantamount to using `t Hooft anomaly matching. In the current case, we wish to highlight how a similar notion of anomaly matching would apply from the 10d theory to the 4d theory \cite{Cordova:2018cvg}. An important ingredient in our specific example will be to identify the (approximate) global symmetries in the 10d action, which eventually match on to the PQ symmetries of the axions. The axion couplings follow from identifying the appropriate symmetries and their mixed anomalies with gauge symmetries. Our treatment highlights an application of global higher-form symmetries to the heterotic case.\footnote{In modern parlance, anomalies refer to `t Hooft anomalies in which only global symmetries participate. When gauge symmetries are involved then often the ``anomaly coefficient" is still physical, and can be cast into a discrete global subgroup, a non-invertible symmetry structure, or a higher-group structure~\cite{Cordova:2018cvg}. We will continue to use the term anomaly to describe mixed global-gauge anomalies.} 

The first step is to identify the origin of U(1) global symmetries in the 10d picture. The 2-form $B$ field has  higher-form global $U(1)$ symmetries in analogy with the electric and magnetic one-form symmetries of the 4d Maxwell theory. The symmetries in the present case are a $2$-form electric and a $6$-form magnetic symmetry. These higher form symmetries reduce to 0-form shift symmetries upon compactification.

We would expect the symmetries above to act as shift symmetries of the two-form $B$ and its dual $B_6$:
\begin{align}
    &B_6 \to B_6 + \Lambda_6 \, ,
    && B \to B + \Lambda_2 \, .
    \label{eq:10dshift}
\end{align}
The gauge-invariant currents for the higher-form global symmetries are $H$ and $H_7$,
\begin{align}
    H &= dB_2 +\frac{1}{16\pi^2}\left (\omega^{\rm grav}_3 - \omega_3\right )\,,
    \\
    H_7 &= dB_6 +\frac{1}{768 \pi^3} X_7 \, ,
\end{align}
where $dX_7 = X_8$ and for the $E_8 \times E_8$ theory $X_8$ is given by: 
\begin{equation}
    \label{eq:8_form}
    X_8= -(\tr _1\mathcal{F}^2+\tr _2\mathcal{F}^2)\tr \mathcal{R}^2+2\left[(\tr _1\mathcal{F}^2)^2+(\tr _2\mathcal{F}^2)^2-\tr _1\mathcal{F}^2\tr _2\mathcal{F}^2\right]\, ,
\end{equation}
where here and throughout the paper we ignore purely gravitational terms which do not affect axion couplings to gauge bosons. For the $SO(32)$ theory we have instead:
\begin{align}
    \label{eq:8_form2}
    X_8= 8 \tr \mathcal{F}^4 - (\tr \mathcal{F}^2)^2 \, .
\end{align}
Each current has an anomalous divergence: 
\begin{align}\label{eq:div_currents}
    & dH = \frac{1}{16\pi^2} \left( \tr \mathcal{F}^2-\tr \mathcal{R}^2 \right)\,, 
    && d H_7= \frac{X_8}{768 \pi^3}\, .
\end{align}

The simplest way to match anomalies between 10d and 4d is to compare the anomaly polynomials (which are $d+2$ forms for a $d$-dimensional QFT).
If we couple the two currents to background gauge fields $V_7, \, V_3$ through the action terms:
\begin{align}
    & \int V_7 \wedge H \, ,
    && \int V_3 \wedge H_7 \, ,
\end{align}
the anomaly polynomials for the electric and the magnetic shift symmetries is the 12-form
\begin{align}
    & \mathcal{I}_{12} = \frac{1}{16\pi^2} dV_7 \wedge \left( \tr \mathcal{F}^2-\tr \mathcal{R}^2 \right) + \frac{1}{768 \pi^3} d V_3  \wedge X_8 \, .
    \label{eq:AP}
\end{align}
We want to match these to the 4d anomaly polynomial obtained upon compactification, which will describe the anomaly of the axion PQ symmetries with the SM gauge group. This will be a 6-form coupling the background gauge field for the $U(1)$ shift symmetry to the low energy gauge groups. 

We can integrate over the compact space and identify the background field strengths associated with various $U(1)$ symmetries in the 4d theory. These background fields we denote $F^{(a)}$ and $F^{(b_i)}$ and are given by
\begin{align}
    &F^{(a)} = \int_K dV_7 \, ,
    &&F^{(b_i)} = \int_{C_i} dV_3 \, .
\end{align}
where $C_i$ are a basis of 2-cycles in the internal manifold $K$. The anomaly polynomial in the 4d theory comes from integrating $\mathcal{I}_{12}$ over~$K$:
\begin{equation}
\begin{aligned}
    \mathcal{I}_6 &= \frac{1}{16\pi^2} \int_K dV_7 \wedge \left( \tr \mathcal{F}^2-\tr \mathcal{R}^2 \right) + \frac{1}{768 \pi^3}  \int_K d V_3  \wedge X_8
    \, .
\end{aligned}
\end{equation}
The first term simply becomes
\begin{align}
     \frac{1}{16\pi^2}F^{(a)} \wedge \left( \tr \mathcal{F}^2-\tr \mathcal{R}^2 \right) \, .
     \label{eq:6-form-anomaly}
\end{align}
To simplify the second term we can expand $dV_3$ in a basis of harmonic 2-forms $\beta_i$ dual to $C_i$ as $ dV_3 = \frac{1}{2\pi} \sum_i F^{(b_i)} \wedge \beta_i$, where $F^{(b_i)}$ has indices along the 4d directions. The second term in the anomaly polynomial then becomes
\begin{align}
    \frac{1}{768 \pi^3}  \int_K d V_3  \wedge X_8 & = \frac{1}{1536 \pi^4} \sum_i F^{(b_i)} \wedge \int_K \beta_i \wedge X_8\,. 
\end{align}
Focusing on the $E_8 \times E_8$ theory we can simplify this expression further down to:
\begin{align}
    \frac{1}{1536 \pi^4} \int \beta_i\wedge X_8
    = \frac{1}{16\pi^2} \left( n^{(1)}_i\tr _1\mathcal{F}^2 + n^{(2)}_i \tr _2\mathcal{F}^2 \right) \, .
\end{align}
where the quantised coefficients $n_i^{(1,2)}$ depend on the background profiles of the gauge fields as:
\begin{equation}
\begin{aligned}\label{eq:model_dep_coeff}
    n^{(1)}_i &= \frac{1}{48\pi^2} \int_K\beta_i\wedge \left( 2 \tr_1\mathcal{F}^2 -  \tr_2\mathcal{F}^2 - \frac12 \tr \mathcal{R}^2 \right) \,,\\
    n^{(2)}_i &= \frac{1}{48\pi^2} \int_K\beta_i\wedge \left( 2 \tr_2\mathcal{F}^2 -  \tr_1 \mathcal{F}^2 - \frac12 \tr \mathcal{R}^2 \right)\,.
\end{aligned}
\end{equation}
Through use of the Bianchi identity~\eqref{eq:bianchi} it can further be shown that $n^{(1)}_i = - n^{(2)}_i \equiv n_i$~\cite{Choi:1985bz}, and the second term in the anomaly polynomial becomes:
\begin{align} 
    \frac{1}{768 \pi^3}  \int_K d V_3  \wedge X_8 & = 
    \frac{1}{16\pi^2} \sum_i n_i F^{(b_i)} \wedge \left( \tr _1 \mathcal{F}^2  - \tr _2 \mathcal{F}^2 \right) \, .
\end{align}
When $\tr_1 \mathcal{F}^2 \neq \tr_2 \mathcal{F}^2$ the $\mathbb{Z}_2$ symmetry which exchanges the two $E_8$ factors is broken by the background. When this is the case some of the $n_i$'s are non-zero and there is a mixed anomaly in 4d.

\section{Axions in heterotic string theory}
\label{sec:heterotic_axions}
In this section we review the origin of axions in heterotic string theory and their couplings to gauge bosons. The different kinds of axions that appear in the heterotic string can be classified as: the model independent axion (which is the integral of $B_6$ over $K$), the model dependent axions arising as the zero modes of $B$ wrapping the $2$-cycles $C_i$ in the compact space, and field theory axions coming from the phases of complex scalars charged under $U(1)$ gauge groups in the effective theory. 

The axions originating from modes of $B$ and $B_6$ inherit their shift symmetries from the higher-form symmetries discussed in the previous section. Their couplings to gauge bosons are set by the  coefficients of the anomaly between the global higher-form symmetry and the gauge symmetry, which were also derived in the previous section. The field theory axions arise in specific models when there are $U(1)$ factors in the low energy gauge group. Often they are heavy as their shift symmetries are not protected by higher-form symmetries, however, in some models they can be light or mix with the other axions in the theory. 

The main result of this section is the 4d Lagrangian, equation~\eqref{eq:MD_lag5}, describing the couplings of all axions to the SM gauge groups for a general embedding of the SM into the $E_8 \times E_8$ theory. This Lagrangian will be used in section~\ref{sec:4d_eft_pheno} to study the IR phenomenology of axions in heterotic string theory. We use a convention that the axion fields are all dimensionless with periodicity $2\pi$ and we neglect the gravitational couplings of the axions.

\subsection{The model independent axion}
The model independent axion ($a$) is the integral of the dual 6-form over the entire internal manifold $K$:
\begin{align}
    a= 2\pi\int_K B_6 \, .
\end{align}
The shift symmetry of the axion is inherited from the 6-form symmetry~\eqref{eq:10dshift} of $B_6$ in the higher dimensional theory
\begin{align}
    a \to a + 2\pi \int_K \Lambda_6 \,.
    \label{eq:MIshift}
\end{align}
It has properties which are largely independent of the details of the compactification and appears in the low energy theory as the pseudo-scalar partner of the dilaton. To determine the action for $B_6$ we dualise the action~\eqref{eq:sugra_10} in terms of $B$ in the standard way, which gives 
\begin{align}
    S_{B_6}= \int d^{10}x \left ( 
    \frac{g_s^2 l_s^4}{4\pi} H_7 \wedge \star H_7
    - \frac{1}{16\pi^2} B_6 \wedge \tr \mathcal{F}^2 \right)
=   \int d^4 x
\left(
    \frac12 f_a^2 da \wedge \star da
    -\frac{a}{16\pi^2} \tr \mathcal{F}^2 
    \right)
    \,,
    \label{eq:B6action}
\end{align}
where we have used the relations $g_s = e^\phi$ and $4\pi\kappa_{10}^2=l_s^8$. This shows that the model-independent axion has a decay constant given by $f_a^2 = \frac{g_s^2 l_s^4}{2\pi \text{Vol}_K}$ and couples to $\tr \mathcal{F}^2$, interacting universally with all gauge bosons. The coefficient of $a \, \tr \mathcal{F}^2$ coupling matches the mixed anomaly coefficient~\eqref{eq:6-form-anomaly} between the 6-form symmetry and the gauge symmetry in the 10d theory.

\subsection{Model dependent axion couplings}
\label{subsec:modeldependent}

The model dependent axions ($b_i$) arise as zero modes of the two-form field wrapping $2$-cycles in the compact space:
\begin{align}
    b_i = 2\pi \int_{C_i} B \, ,
\end{align}
or equivalently as coefficients of the expansion of $B$ in terms of harmonic two-forms:
\begin{equation}
    B=\frac{1}{2\pi}\sum_{i=1}^{h_{11}} b_i \beta_i \,.
    \label{eq:MDaxions}
\end{equation}
The shift symmetry of $b_i$ is inherited from the 2-form symmetry~\eqref{eq:10dshift} of $B$ restricted to the cycle $C_i$
\begin{align}
    b_i \to b_i + 2\pi \int_{C_i} \Lambda_2 \,.
    \label{eq:MDshift}
\end{align}
The number of model-dependent axions and their couplings depend on topological data only, that is the number of 2-cycles (the second Betti number, $h_{11}$) and the fluxes through them. Their decay constants and masses depend additionally on dynamical data such as the volume of the cycles. 

The couplings of the model dependent axions are set by the coupling of $B$ which cancels the 10d anomalies via the Green-Schwarz mechanism~\cite{Green:1984sg}, and are different for each $E_8$ factor~\cite{Choi:1985bz,Derendinger:1985cv}. The relevant term in the action is:
\begin{equation}\label{eq:10D_GS}
    -\frac{1}{768 \pi^3} \int B\wedge X_8\, ,
\end{equation}
where $X_8$ is defined in equation~\eqref{eq:8_form}. Integrating over the internal manifold and following the same logic as in section~\ref{sec:anomalymatching} we find the couplings
\begin{align}  \label{eq:axion_coupling_heterotic}
    &S_{\rm MD} = - \frac{1}{16\pi^2}\sum_i \int_{M_4} n_i b_i  
    \left(\tr _1\mathcal{F}^2 - \tr _2\mathcal{F}^2 \right) \, ,
\end{align}
with $n_i$ defined in equation~\eqref{eq:model_dep_coeff}.

\subsection{$U(1)$ factors and field theory axions}

\label{sec:anomalous_U(1)}
If there are $U(1)$ factors in the gauge group after compactification they can modify the low energy axion spectrum. In these scenarios there will generically be additional complex scalars $\phi_j$ charged under the $U(1)$'s which are also singlets under the other parts of the 4d gauge group. These come from extra-dimensional components of 10d gauge fields upon dimensional reduction. If these scalars get a vev, $\phi_j = f_{c_j} e^{i q_{c_j} c_j (x)} /\sqrt{2}$, the phase $c_j$ can act as an axion. Under these $U(1)$'s the various axions transform as
\begin{align}
   &a (x) \rightarrow a (x) + \sum_n \lambda^{(n)} (x) q_{a}^{(n)} \, ,
   \nonumber
   \\
   & b_i(x)\rightarrow b_i(x) + \sum_n\lambda^{(n)}(x) q_{i}^{(n)} \, ,
   \\
   & c_j(x) \rightarrow c_j(x) + \sum_n \lambda^{(n)} (x) q_{c_j}^{(n)} \, ,
   \nonumber
\end{align}
where $n$ indexes the $U(1)$ factors in the effective theory, the $q$'s are the charges of each axion and the $\lambda^{(n)}(x)$ are the gauge transformation parameters.  

If these $U(1)$'s are not anomalous from the perspective of the 4d theory, then it is possible that none of the axions will transform under the $U(1)$, in which case it will remain unbroken. If these $U(1)$ gauge groups have an anomalous fermion spectrum the anomaly is cancelled by a 4d version of the Green-Schwarz mechanism. Furthermore, there will be a $D$-term potential in the effective theory which means that at least one complex scalar will get a vev~\cite{Dine:1987xk}, and so there will always be an associated $c_j$ axion charged under this $U(1)$\footnote{For example, taking a line bundle in the internal dimensions with structure group $S(U(1)^5)$ leaves an unbroken $SU(5)\times U(1)^4$ symmetry~\cite{Blumenhagen:2005ga,Anderson:2011ns}.}. One linear combination of higher-form axions $a$ and $b_i$, together with the phases $c_i$ will be eaten to make the gauge boson massive, while the orthogonal linear combinations remain as a light axions at low energies. The concrete linear combination depends on the $U(1)$ charges, $q_i$.

Note that the $c_j$'s never add any light axions on top of those coming from $B$. For each broken $U(1)$ we can rotate to a basis where there is one linear combination, which we denote $c_n$, which transforms under the gauge symmetry:
\begin{align}
    & c_n(x) \rightarrow c_n(x) + \lambda^{(n)} (x) q_{c}^{(n)} \, .
\end{align}
The orthogonal linear combinations of $c_j$ do not shift under the $U(1)$ and will get a gauge invariant perturbative mass. Each $c_n$ can further mix with the axions coming from the $B$ field, with one combination of $a, b_i, c_n$ being eaten and the other combinations remaining as axions at low energies. The field theory axions will in general couple separately to each $E_8$ factor, with couplings described by the anomaly coefficients $m^{(1)}_n, \, m^{(2)}_n$:
\begin{align}
    S_{\rm c, int} = \frac{1}{16 \pi^2}  \sum_n \int_{M_4} c_n
    \left( m^{(1)}_n  \tr_1 \mathcal{F}^2 + m^{(2)}_n \tr_2 \mathcal{F}^2 \right) \, .
    \label{eq:FTaxioncouplings}
\end{align}

\subsection{All together}
We collect the results described in this section above to write the full Lagrangian describing the axions in the low energy theory. The gauge invariant kinetic terms for $a$ and $c_n$ are diagonal and given by:
\begin{align}
    \mathcal{L}_{a,\rm{kin}} =\frac{1}{2} f_a^2
    \left( \partial_\mu a + \sum_n q_{a}^{(n)} A^{(n)}_\mu\right)^2
    \, ,     \\
     \mathcal{L}_{c,\rm{kin}} =\frac{1}{2} \sum_n f_{c_n}^2\left(\partial_\mu c_n + q_{c}^{(n)}  A^{(n)}_\mu \right)^2
     \, .
\end{align}
The kinetic terms for the $b_i$ are off diagonal, however, with a mixing matrix given by~\cite{Svrcek:2006yi}:
\begin{align}
    \gamma_{ij} = \int_K \beta_i \wedge \star \beta_j \, .
    \label{eq:mixingmatrix}
\end{align}
The kinetic terms for the $b_i$ then read
\begin{align}
     \frac{1}{2} \sum_{ij} \gamma_{ij} f_{b_i} f_{b_j} g^{\mu\nu} \left( \partial_\mu b_i + \sum_m  q_{i}^{(m)} A^{(m)}_\mu\right)
    \left( \partial_\nu b_j + \sum_n  q_{i}^{(n)} A^{(n)}_\nu \right) \, .
\end{align}
These kinetic terms can be made diagonal by a combination of a rescaling  $f_{b_i} \to \tilde f_{b_i}$ and a rotation $b_i \to \tilde b_i = R^T_{ij} b_j$, such that $ f_{b_i} f_{b_j} \left( R^T \gamma R \right)_{ij} = \tilde f_{b_i} \tilde f_{b_j} \delta_{ij}$. In this basis the kinetic terms for the model dependent axions read
\begin{align}
     \mathcal{L}_{b,\rm{kin}} =\frac{1}{2} \sum_{i} \tilde f_{b_i}^2 \left( \partial_\mu \tilde b_i + \sum_n \tilde q_{i}^{(n)}A^{(n)}_\mu\right)^2  \, ,
\end{align}
where $\tilde q_{i}^{(n)} = R_{ik} q_{k}^{(n)}$. The anomalous couplings $n_i$ are also rotated by this change of basis, $\tilde n_i = R_{ik} n_k$.

In the diagonal basis the kinetic terms in the 4d Lagrangian describing the axions become:
\begin{align}
    \mathcal{L}_{4,\rm{kin}} =&
    \frac{f_a^2}{2} 
     \left( \partial_\mu a + \sum_n q_{a}^{(n)} A^{(n)}_\mu\right)^2
     +  \sum_{i=1}^{h_{11}} \frac{\tilde f_{b_i}^2}{2}  \left( \partial_\mu \tilde b_i + \sum_n \tilde q_{i}^{(n)}A^{(n)}_\mu\right)^2  
     + \sum_n \frac{f_{c_n}^2}{2}\left(\partial_\mu c_n + q_{c}^{(n)}  A^{(n)}_\mu \right)^2\,.
\end{align}
The interaction terms are:
\begin{align}
    \mathcal{L}_{4,\rm{axion}} =&
     \frac{a}{16\pi^2}
     (\tr_1 \mathcal{F}^2 + \tr_2 \mathcal{F}^2)
     +\frac{1}{16 \pi^2}  \sum_n  c_n
    \left( m^{(1)}_n  \tr_1 \mathcal{F}^2 + m^{(2)}_n \tr_2 \mathcal{F}^2 \right) 
     +\frac{1}{16\pi^2} \sum_{i=1}^{h_{11}} \tilde n_i \tilde b_i  
    \left(\tr _1\mathcal{F}^2 - \tr _2\mathcal{F}^2 \right) \,.
    \label{eq:MD_lag3}
\end{align}
Using equation~\eqref{eq:embeddinge8} we get,
\begin{align}
    \mathcal{L}_{4,\rm{axion}} =&
     \frac{1}{16\pi^2}
     \left( (a + c_n m^{(1)}_n + \tilde n_i \tilde b_i)
     \sum_l
     k^{(1)}_l \tr ({F_l^{(1)}})^2 
     +
     (a + c_n m^{(2)}_n - \tilde n_i \tilde b_i)
    \sum_l k^{(2)}_l \tr ({F_l^{(2)}})^2
    \right) \, ,
    \label{eq:MD_lag4}
\end{align}
where the sums over axions are left implicit. We can extract the couplings to the SM gauge groups using equation~\eqref{eq:embed10dto4d},
\begin{align}
    \mathcal{L}_{4,\rm{axion}} 
    &=
     \frac{1}{16\pi^2}
     \left(
     \frac{1}{k_1^{(1)}+k_1^{(2)}}
     \left[
     (k_1^{(1)})^2 
      (a + c_n m^{(1)}_n + \tilde n_i \tilde b_i)
      +
     (k_1^{(2)})^2  (a + c_n m^{(2)}_n - \tilde n_i \tilde b_i)
     \right]
     F_Y\wedge F_Y
     \right.\nonumber \\&
     \qquad 
     +
     \frac{1}{k_2^{(1)}+k_2^{(2)}}
     \left[
     (k_2^{(1)})^2 
      (a + c_n m^{(1)}_n + \tilde n_i \tilde b_i)
      +
     (k_2^{(2)})^2  (a + c_n m^{(2)}_n - \tilde n_i \tilde b_i)
     \right]
     \tr( W\wedge W)
   \nonumber \\&
     \qquad 
     +
     \left.
          \frac{1}{k_3^{(1)}+k_3^{(2)}}
     \left[
     (k_3^{(1)})^2 
      (a + c_n m^{(1)}_n + \tilde n_i \tilde b_i)
      +
     (k_3^{(2)})^2 (a + c_n m^{(2)}_n - \tilde n_i \tilde b_i)
     \right]
     \tr(G\wedge G)
    \right) + \ldots
    \label{eq:MD_lag5}
\end{align}
The general situation is that the axions coupled to each of the SM gauge groups are combinations of $\varphi_1$ and $\varphi_2$ where:
\begin{equation}
\begin{aligned}
    \varphi_1 &= a + c_n m^{(1)}_n + \tilde n_i \tilde b_i \, ,           \\
    \varphi_2 &=  a + c_n m^{(2)}_n - \tilde n_i \tilde b_i\, .
    \label{eq:varphis}
\end{aligned}
\end{equation}
Here we note that $\tilde n_i \tilde b_i = n_i b_i$, so that under a $2\pi$ shift of $b_i$, $\varphi_1 \to \varphi_1 + 2\pi n_i , \, \varphi_2 \to \varphi_2 - 2\pi n_i  $, maintaining the axion periodicity.

\section{4D EFT and phenomenology of heterotic string axions}\label{sec:4d_eft_pheno}

In the previous section we obtained the 4d Lagrangian~\eqref{eq:MD_lag5} describing the axions and their couplings to gauge bosons. The Lagrangian was derived by matching the 4d axion couplings to anomaly coefficients in the 10d theory and rotating to a basis where the kinetic terms of the axions are diagonal. The remaining unknown (or model-dependent) quantities are contributions to the axion potential, which we parameterise by a scale $\Lambda$ where relevant. These contributions are discussed further in appendix~\ref{app:shifts} and possible corrections due to twisted states are discussed in appendix~\ref{app:twisted}.

The goal of this section is to show that the presence or not of ALPs which evade the bound~\eqref{eq:bound_coupling}, which we refer to as a `light ALP',  depends only on how the SM gauge group is embedded into the UV gauge group. In section~\ref{sec:SO32} and~\ref{sec:heterotic_QCD} we show that when the SM is embedded into a single gauge group (as is the case for the $SO(32)$ theory or when the SM comes entirely from one $E_8$ factor) all axions satisfy the bound~\eqref{eq:bound_coupling}. In other scenarios $\varphi_2$ can be a light ALP coupled to photons if either hypercharge or QCD comes from a combination of gauge groups from both $E_8$ factors. This case is discussed in section~\ref{subsec:heterotic_ALP}. 
There is also the possibility, discussed in section~\ref{subsec:kinetic}, where kinetic mixing of the photon can induce a photon coupling of a light ALP which does not get a mass from QCD, although the ALP-photon coupling is suppressed in this case. The consequences of each case for various experiments is summarised in section~\ref{subsec:experiments}.

\subsection{The $E_8$ QCD axion}
\label{sec:heterotic_QCD}

In this section we study the axion couplings in compactifications where the SM is embedded into one $E_8$ factor:
\begin{equation}
    E_8 \supset \mathbb{G} \supset SU(3)_C\times SU(2)_w\times U(1)_Y\,.
\end{equation}
This corresponds to the case where $k^{(2)}_1 = k^{(2)}_2= k^{(2)}_3 = 0$. This occurs in many known models, for example in situations where we embed the spin connection into a single $E_8$. In this case the hidden sector is an $E_8$ gauge group which may be further broken by Wilson lines or Higgs fields. In orbifold models, or models where the spin connection/general vector bundle is embedded into both $E_8$'s, the second $E_8$ will also be broken to a subgroup before Wilson lines are taken into account. The hidden sector will generate contributions to the potential either from UV stringy instantons or confining gauge groups. In this section we don't assume a particular form of the hidden sector. We well see that in this scenario, the bound in equation~\eqref{eq:bound_coupling} is obeyed regardless of the details of the compact space.

From the perspective of an IR observer, the only relevance of any additional gauge groups (from either $E_8$ factor) on top of the SM gauge group is their contributions to the axion potential. For example, if the hidden sector is an unbroken $E_8$ then the linear combination $\varphi_2$ will receive a large mass from the confinement of $E_8$~\cite{Choi:1985je}. Gauge sectors from either $E_8$ will combine with contributions to the potential from worldsheet instantons~\cite{Wen:1985jz} to lead to an effective potential $V$ for the axion fields. We discuss possible contributions to the potential in Appendix \ref{app:shifts}.

The relevant Lagrangian terms at low energies are then
\begin{align}\label{eq:eff_Lag_E8_axion}
    \mathcal{L} &= \frac{\varphi_1 }{8\pi }
    \left(\alpha_1k_1^{(1)} F_Y \tilde F_Y +\alpha_2k_2^{(1)} \tr W \tilde W
    + \alpha_3 k_3^{(1)} \tr G \tilde G\right ) + V(\varphi_i) \, ,
\end{align}
where $F_Y$, $W$, and $G$ are the field strengths for the SM gauge groups $U(1)_Y$, $SU(2)_L$, and $SU(3)_C$, respectively. $\varphi_1$ has already been defined in equation~\eqref{eq:varphis} and we denote all other axions by $\varphi_i$, $i\geq 2$ for some orthonormal basis (not necessarily the mass basis). Below the QCD scale this leads to the Lagrangian for the QCD axion coupled to QED
\begin{align}
    \mathcal{L} &= \frac{\varphi_1}{8\pi } 
    \left( \frac{E}{N}  - 1.92 \right) \alpha_\EM F_\EM \tilde F_\EM  + V_{\rm QCD} (\varphi_1)  + V(\varphi_i) \, ,
    \nonumber
    \\
    \frac{E}{N} &=\frac{k_2^{(1)}+k_1^{(1)}}{k_3^{(1)}}\, ,
    \label{eq:EN}
\end{align}
where $V_{\rm QCD}$ is the QCD potential given in equation~\eqref{eq:QCDcoupling}. In this case, the constant $C$ in equation~\eqref{eq:bound_coupling} takes the same values as in 4d unified theories (given in equation~\eqref{eq:const_qcd}). We note that the mass matrix derived from $V$ must have a zero eigenvalue in a direction which overlaps with $\varphi_1$ in order for the theory to solve the strong CP problem.

To obtain the experimentally relevant axion couplings we need to rotate to a basis of mass eigenstates. While the details of this will depend on the specific model, the point is that $\varphi_1$ can be expanded in terms of mass eigenstates $\tilde\varphi_i$  (which we take to be dimensionful with periodicity $f_i$) as
\begin{align}
    \varphi_1 = \sum_i R_{1, i} \frac{\tilde \varphi_i}{f_i}\,,
\end{align}
where $R_{ij}$ is the matrix which rotates to the mass basis. This is precisely the same result as was found for 4d unified theories in~\cite{Agrawal:2022lsp}. The main implication is that the axion $\tilde \varphi_i$ couples to both electromagnetism and QCD equally, with coupling $R_{1, i} /f_i$. Axions lying the left of the QCD line are not possible, as any coupling to photons comes correlated with a QCD coupling which generates a mass (in addition, possibly, to other mass contributions coming from $V$). For light axions which couple through mass mixing, the coupling will be suppressed by a small parameter:
\begin{equation}
    R_{1, i} \sim m_i^2/m_{\rm QCD}^2\,,
\end{equation}
where $m_i$ is the light axion mass and $m_{\rm QCD}$ is the mass of an axion which falls on the QCD line~\cite{Agrawal:2022lsp}, defined as parameters which saturate the bound~\eqref{eq:bound_coupling} with $C$ given in equation~\eqref{eq:const_qcd}. There is no quality problem in this model if the mass matrix from $V$ has a vanishing (or sufficiently small) eigenvalue, as may be expected for a potential generated by instantons. In this case there is a QCD axion which saturates the bound~\eqref{eq:bound_coupling} and solves the strong CP problem. 

The value of $E/N$ is the only undetermined quantity which determines the position of the QCD line in a given model. In all unified theories which have been considered, $E/N$ takes the value $E/N=8/3$, which is also the typical expectation in heterotic models. However the possibilities are more general in this case, with the only requirements being that the couplings satisfy $k_1 \alpha_1 =  k_2 \alpha_2 =  k_3 \alpha_3$ at the compactification scale and $k_1 \geq 1$ for the existence of the lepton singlet. It is difficult to obtain a precise upper bound on $E/N$ as the running of the SM gauge couplings depend on the matter content between the electroweak and GUT scales, however, given some assumptions the range of possible values for the factor $E/N - 1.92$ are:
\begin{align}
   0.08 <  \frac{E}{N} -1.92 \lesssim  30 \, \, \, .
   \label{eq:ENbound}
\end{align}
The lower bound comes from choosing $k_1 = k_2 = k_3 = 1$, while the upper bound comes from taking $k_2 = 3, \, k_3 = 1, \, k_1 = 57/2$. The first choice maximises the cancellation between the quantised and irrational pieces, while the second choice maximises $E/N$ while satisfying the assumptions of ref.~\cite{Font:1990uw} (this assumes MSSM running up to the GUT scale, which is taken to be $ \geq 10^{15}$ GeV to avoid rapid proton decay). Different assumptions about the running of couplings could change the upper bound in equation~\eqref{eq:ENbound}, but the general expectation is that $E/N \sim \mathcal{O}(1)$ and it is difficult to enhance the axion coupling by a significant amount by appealing to a large $E/N$~\cite{Dienes:1995sq}. Typical strategies to increases $E/N$, such as clockwork models~\cite{Farina:2016tgd, Agrawal:2017cmd} are not available in heterotic theories.

\subsubsection{The $SO(32)$ case}
\label{sec:SO32}
For the sake of completeness, let us mention the case of heterotic $SO(32)$. In this case, the model independent axion still couples to $\tr (\mathcal{F}^2)$ as described by the action~\eqref{eq:B6action}. After using the appropriate $SO(32)$ relations for the traces and the Bianchi identity, the model dependent axion couplings from the GS term are: 
\begin{equation}
    \label{eq:couplings_model_dep_SO32}
    \frac{1}{768 \pi^3} \int B \wedge X_8=\frac{1}{1536\pi^4}\sum_i \int_K b_i \beta_i \wedge \left( 8 \tr \mathcal{F}^4 - (\tr \mathcal{F}^2)^2  \right) \,.
\end{equation}
The $\tr \mathcal{F}^4$ does not lead to couplings of axions to SM gauge bosons in 4d. Such a coupling contribution would come from traces where two generators are those of background fields, while the other two are SM generators. The SM generators must commute with the background gauge fields (and be distinct from them) and the trace can be shown to vanish in this case.

The equation above implies that, just as in 4d GUTs, there is a single linear combination of axions which couples to the gauge bosons which survive the breaking by embedding the spin connection and  turning on Wilson lines. These gauge bosons, which we have denoted in compact notation by $\tr \mathcal{F}^2$, correspond to the generators in $SO(32)$ which commute with the structure group of the vector bundle, and the Wilson line. Such a linear combination includes the model-independent axion together with $b_i$:
\begin{equation}
    \varphi=a+\sum_i n' b_i\,,
\end{equation}
for some quantised coefficients $n'$. This linear combination can couple to photons and receives a mass from QCD instantons. Additional instantons in $SO(32)$ or other non-perturbative  stringy effects may increase the axion mass, and reintroduce a quality problem. 

\subsection{Light ALPs from non-standard embedding}
\label{subsec:heterotic_ALP}
If the Standard Model is embedded in a more involved way into $E_8\times E_8$ then it may be that there are ALPs with photon couplings but no gluon coupling. This will be the case if the SM is embedded into $E_8 \times E_8$ in such a manner so that the axion which couples to QCD is different from the axion coupling to hypercharge. This will be the case if the ratios $E/N$ are different for the axions coupling to each different $E_8$:
\begin{align}
    \frac{k_1^{(1)} + k_2^{(1)}}{k_3^{(1)}} \neq \frac{k_1^{(2)} + k_2^{(2)}}{k_3^{(2)}}\,.
\end{align}
One possibility is if QCD and $SU(2)_w$ are embedded into the first $E_8$ and hypercharge receives a contribution from a $U(1)$ subgroup of the second $E_8$ factor, $k_3^{(2)} = k_2^{(2)} = 0$. There are many orbifold models which have been considered where there are multiple $U(1)$ factors in the low energy theory, coming from both $E_8$ groups in the UV~\cite{Ibanez:1987sn,Font:1989aj,Choi:2006qj,Choi:2009jt}. These models also have twisted states with $U(1)$ charges from each $E_8$, which could potentially higgs combinations of the $U(1)$s to a diagonal subgroup. This would leave a $U(1)$ gauge group in the IR which is a linear combination of generators from both $E_8$s, and may be identified with hypercharge in some models. It is unknown to the authors if a realistic model has been constructed in this way, since these models typically embed the SM in an $SU(5)$ subgroup.  However, since the necessary ingredients are present in orbifold constructions we consider this possibility here.

Comparing to 4d unified theories this situation looks very much like flipped $SU(5)$. Starting with equation~\eqref{eq:MD_lag5}, below the compactification scale we can write the relevant interaction Lagrangian for $\varphi_1, \, \varphi_2$ as: 
\begin{equation}\label{eq:eff_Lag_non-standard_embedding}
    \mathcal{L}= \frac{\varphi_1}{8\pi} \left[ k^{(1)}_3 \alpha_3 \tr G\tilde{G} +   \left( \frac{(k^{(1)}_1)^2}{k^{(1)}_1 + k^{(2)}_1} + k^{(1)}_2 \right)\alpha_\EM F_\EM  \tilde{F}_\EM \right] 
    + \frac{\varphi_2}{8\pi} \left( \frac{(k^{(2)}_1)^2}{k^{(1)}_1 + k^{(2)}_1}  \right)\alpha_\EM F_\EM  \tilde{F}_\EM 
    + V_{\eff} (\varphi_i)\, ,
\end{equation}
where the embedding levels $ k^{(1)}_3, \,  k^{(1)}_2, \,  k^{(1)}_1$ are the embedding levels of the SM gauge groups in the first $E_8$ and $k^{(2)}_1$ is the embedding level of hypercharge into the second $E_8$.

There are several effects which contribute to the axion potential in any given model (see Appendix \ref{app:shifts}). There will always be one contribution which generates a potential specifically for the linear combination $\varphi_2$ from gauge instantons from the second $E_8$ group. We parametrise this piece as:
\begin{equation}
    V_{\eff} \sim - \Lambda^4\cos (\varphi_2) \, .
    \label{eq:phi2potential}
\end{equation}
The unknown coefficient $\Lambda$ is then what determines the smallest possible mass for an axion coupled to photons. $\Lambda$ is set by the hidden sector confinement scale or the action of constrained UV instantons, including possible chiral suppression. Therefore the possible values of $\Lambda$ are model dependent and can span many orders of magnitude.

The mass scale set by the potential~\eqref{eq:phi2potential} is $m_2 = \Lambda^2 /f_2$. If other contributions to the axion potential are large, then there may be no axion with mass equal to $m_2$, however, $m_2$ determines the smallest possible mass for an axion coupled to photons. Combinations of axions $\varphi_i$ (for $i \geq 3$) can couple to QED through mass mixing with $\varphi_2$, but will have a ratio of photon coupling to mass which satisfies
\begin{equation}
    \frac{g_{\varphi_i \gamma\gamma}}{m_{\varphi_i}}\lesssim \frac{1}{\Lambda^2}\,.
    \label{eq:ALPratio}
\end{equation}
This result is analogous to the statement for QCD with the QCD scale replaced by the unknown scale $\Lambda$. For example, if $\varphi_2$ is a mass eigenstate, then axions lighter than $m_2$ will have photon couplings which are suppressed by the ratio :
\begin{align}
    \frac{g_{ \varphi_i \gamma \gamma}}{g_{ \varphi_2 \gamma \gamma}} \propto \frac{m_i^2}{m_2^2} \, .
\end{align}
Heavy axions which mix can have photon couplings $g_{ \varphi_i \gamma \gamma} \sim \mathcal{O}(1) / f_i$, however will still satisfy the ratio~\eqref{eq:ALPratio}. If $\Lambda$ is exponentially small, as may be the case if there are no confining groups in the hidden sector, then there remains the possibility that there are multiple axions the left of the QCD line with observable photon couplings.

This class of models allows for the possibility of a large number of light ALPs with large photon couplings. We note that this can only occur if the scale $\Lambda$ associated with the potential generated from the second $E_8$ factor is much smaller than the QCD scale. Other axions generically have suppressed photon couplings unless there is a coincidence of scales such that $m_i \sim m_2$ for some axion(s) $a_i$ (see \cite{Gavela:2023tzu} for this kind of scenarios with multiple axions with similar $g_a/m_a$). 
\\

\subsection{Light ALPs from $U(1)$ Kinetic mixing}
\label{subsec:kinetic}

An alternative possibility that leads to light ALPs is if $U(1)_Y$ is embedded into a single $E_8$ factor but mixes with a $U(1)'$ gauge boson from the second $E_8$ factor. This requires matter charged under both $E_8$'s, which is present in orbifold models. The ALP coupling to photons is then suppressed by the kinetic mixing parameter making it difficult to observe. If such kinetic mixing can occur, as shown in~\cite{Agrawal:2022lsp}, the coupling of the second axion $\varphi_2 = a_2/f_2$ to the canonically normalised photon will be
\begin{equation}
    \frac{\epsilon^2 \alpha_2}{8\pi}\frac{a_2}{f_2} F_{em}\tilde{F}_{em} \,,
\end{equation}
where $ \alpha_2$ is the coupling of the second $U(1)$. The general expectation for the kinetic mixing parameter is that $\epsilon \sim 10^{-4} - 10^{-2}$~\cite{Goodsell:2011wn}, depending on the model. Given the general expectation that decay constants in heterotic string theory are large~\cite{Svrcek:2006yi} this could make the ALP coming from kinetic mixing very difficult to observe. 
Despite this, the mass of $a_2$ may be determined solely by instanton contributions which are exponentially small and so may be almost arbitrarily light. 

If there is a way for the decay constant $f_2$ to be made small then this can lead to a light ALP which may be observable in experiments. For example, if we take $\epsilon \sim 10^{-2}$, \, $\alpha_2 \sim \alpha_\EM \sim 10^{-2}$, for a decay constant $f_a \sim 10^{6}$~GeV it is possible to get a photon coupling $g_{a \gamma \gamma} \sim 10^{-12}$~GeV$^{-1}$ which can be observed in Chandra observations~\cite{Wouters:2013hua, Marsh:2017yvc,Reynolds:2019uqt, Reynes:2021bpe}. As we expect the axion decay constant to be set by the size of internal cycles on the compactification manifold, such a low decay constant requires the size of some cycles to be much larger than the generic expectation. For such a small decay constant axions which couple to electrons may run into bounds from stellar cooling~\cite{Capozzi:2020cbu}, so any model which realises an ALP through kinetic mixing must also avoid generating a coupling of the axion to electrons.

\begin{figure}[t]
    \centering
    \includegraphics[scale=0.4]{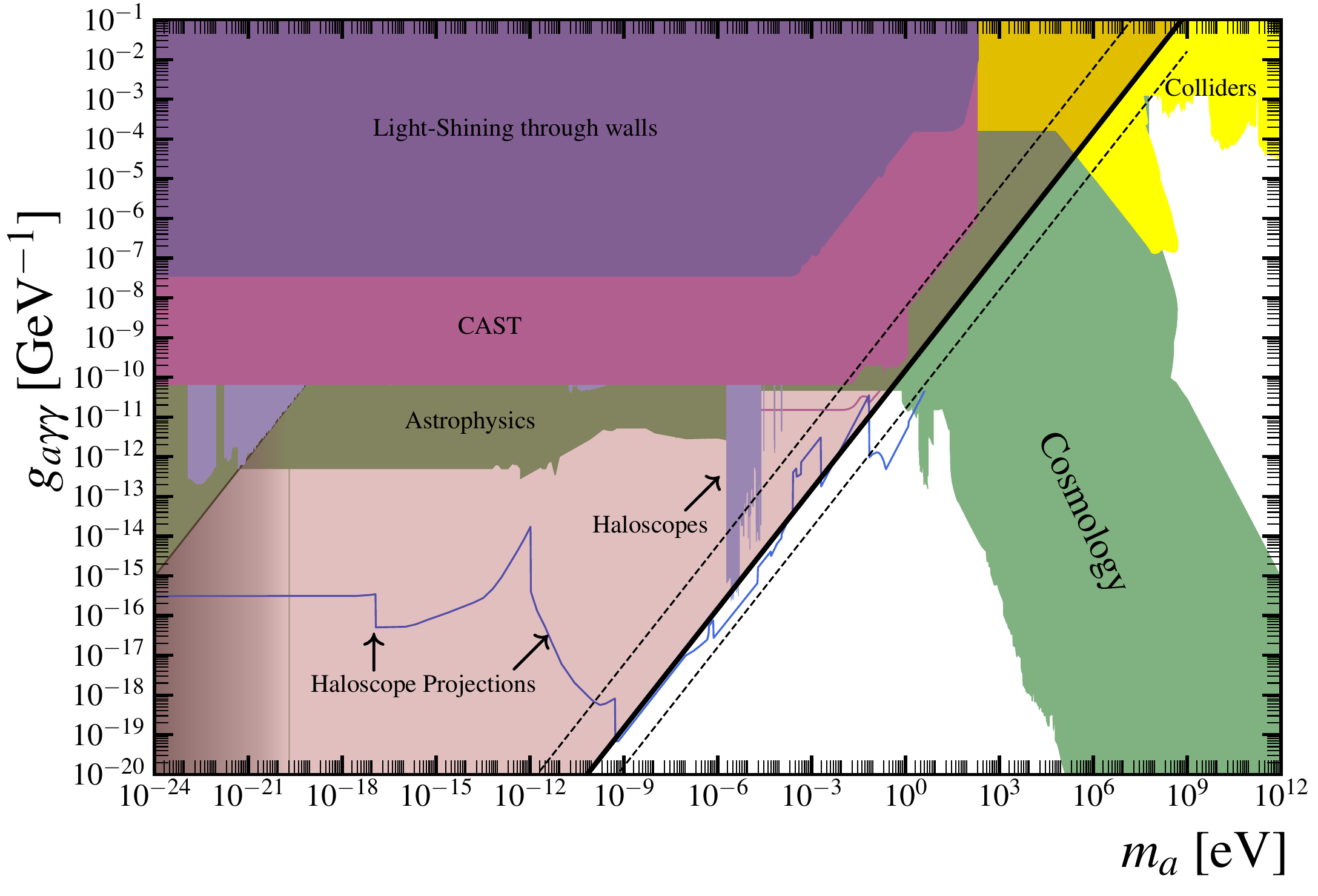}
    \caption{Plot summarising the allowed parameter space in heterotic models and the dominant experimental constraints. The solid black line represents the QCD axion line for the canonical case of $E/N=8/3$, while the dashed lines show the QCD line for $E/N = 2$ and $E/N = 63/2$. The light red shading on the upper left half of the plot indicates the region that is not populated in models where the UV gauge group is $SO(32)$ or the SM is embedded into a single $E_8$ factor. The yellow region shows constraints from colliders, the green region the constraints from cosmological and astrophysical observations, the purple region shows light-shining through walls experiments, and the pink region is the CAST bound. The light blue shaded region shows the current haloscope bounds while the light blue line shows the projected reach. We refer the reader to ref.~\cite{Antel:2023hkf} for a detailed discussion of the experimental constraints. Plot adapted from~\cite{AxionLimits}.}
    \label{fig:ParameterSpace}
\end{figure}

\section{Experimental Implications}\label{subsec:experiments}

In this section we summarise the axion phenomenology of heterotic models and the implications for experiments. In the case of embedding the SM into a single $E_8$ factor we have seen that the QCD axion line defines a boundary in the possible values of $(\ga,m_a)$, see figure~\ref{fig:ParameterSpace}. This is the same result as for 4d GUTs. There is nevertheless the possibility of having $E/N$ be different from the canonical value $E/N=8/3$ which shifts the precise location of the QCD line from the expected GUT value. This allows for the possibility of having parametrically lighter axions than in 4d GUTs if $E/N$ is large, although we are not aware of models which achieve this. Given the restrictions on the SM embeddings, this also comes at the cost of having a non-standard value of $\sin^2\theta_w$ at the compactification scale which requires the running of the gauge couplings to be significantly different compared to the SM. Also, $E/N=2$ is possible, which reduces the coupling to photons by almost an order of magnitude. If one wants to avoid fractional charges, then the weak mixing angle at the GUT scale takes a non-standard value and this may require a non-standard running of gauge couplings.

Any hint of a light ALP would have significant implications for heterotic string models. As discussed above, the only ways this can occur is if hypercharge is embedded into both $E_8$ factors or the photon kinetically mixes with a dark photon. In both cases the ALP gets a mass from instantons or confining sectors in the second $E_8$, which in some cases can be exponentially small. For the case of kinetic mixing the ALP has a photon coupling which is suppressed by the small mixing parameter $\epsilon$, so will in general be out of reach of experiments unless the decay constant is much smaller than the typically expected values, corresponding to a cycle of the internal manifold being atypically large. 

In the near future many experiments will be looking for axions, with most of the parameter space they are sensitive to being above the QCD line. These experiments will therefore be sensitive to the different embeddings of the SM into heterotic string theory. For example, several proposed haloscopes will search for axions with masses ranging from $10^{-6}$ eV  to $O$(1) eV~\cite{Marsh:2018dlj, Lawson:2019brd, Beurthey:2020yuq,Schutte-Engel:2021bqm,Aja:2022csb,ALPHA:2022rxj,ADMX:2023rsk,DeMiguel:2023nmz,Ahyoune:2023gfw,Alesini:2023qed,BREAD:2023xhc}. In the ultralight regime, $10^{-21} \lesssim m_a /\text{eV} \lesssim 10^{-11}$, DMRadio will cover large regions of the parameter space above the QCD line~\cite{DMRadio:2022pkf}, and there are new proposals which will be sensitive to the same region~\cite{Bourhill:2022alm,Oshima:2023csb,Friel:2024shg,Kalia:2024eml}. Finally, there are also other ideas that propose to use polarimetry~\cite{Liu:2021zlt} to search for the imprint of an ultralight axion coupled to photons with masses as light as $m_a\sim 10^{-21}$~eV~\cite{Gan:2023swl}.

A possible hint of a light axion is the recently reported cosmic birefringence signal~\cite{Minami:2020odp,Eskilt:2022cff,Nakai:2023zdr,Diego-Palazuelos:2023mpy,Cosmoglobe:2023pgf}. The signal can be explained by an ALP $b$ with photon coupling $g_{b \gamma \gamma} = \alpha_\EM /f_b$ whose background value changes by an amount $\delta b \sim \pi f_b /\sqrt{3}$ between the time of CMB and today~\cite{Carroll:1989vb, Harari:1992ea, Lue:1998mq}. This requires it to have a mass between the hubble scales at CMB and today: 
\begin{align}
    10^{-31} \lesssim \frac{m_b}{\text{eV}} \lesssim  10^{-28}   \, .
    \label{eq:massrange}
\end{align}
Confirmation of this signal would indicate that if the SM UV completes to a heterotic string model, it must be a model where the UV gauge group is $E_8 \times E_8$ and the SM is embedded in a non-trivial way into both $E_8$ groups. 
As ALPs from kinetic mixing have photon couplings which are suppressed by the mixing parameter $\epsilon$ ($g_{b \gamma \gamma} \sim \epsilon^2 \alpha_\EM /f_b$) they are unable to explain the birefringence signal as $\delta b \leq 2\pi f_b$. The only way to explain the signal in heterotic string theory is if the SM is embedded non-trivially into both $E_8$ factors and the axion potential is sufficiently suppressed to lead to an ALP $b$ which falls in the range~\eqref{eq:massrange}.

For larger masses, $10^{-19}\text{ eV}\lesssim m_a \lesssim 10^{-22}$ eV, the axion can still generate an interesting birefringence signal. This case requires the axion to be the dominant part of the DM to have a sizeable effect, and the signal is an oscillation of the CMB linear polarisation \cite{Fedderke:2019ajk} which has been recently searched for at different observatories \cite{BICEPKeck:2021sbt,SPT-3G:2022ods,POLARBEAR:2023ric}. A similar search using the Crab Nebula as a source has been performed \cite{POLARBEAR:2024vel}. We emphasise that a positive signal in these observatories would also rule out the heterotic string compactifications in which the SM comes from a single $E_8$ and could only be accommodated in models with a non-standard embedding of the SM, such as those discussed in section~\ref{subsec:heterotic_ALP}.

\section{Conclusion}
Despite the large number of possible 4d EFTs describing heterotic string compactifications, there are only two consistent supersymmetric gauge theories in 10d -- $E_8 \times E_8$ and $SO(32)$. In this work we have shown that the axion couplings to gauge bosons in these models are highly restricted by the simplicity of the UV gauge group. The axion shift symmetry derives from a higher-form symmetry in the 10d theory and the axion couplings to gauge bosons are inherited from the mixed anomaly between the higher-form symmetry and the gauge symmetry. We derived these couplings by matching anomalies from the 10d theory to the 4d theory. One of the main implications of our work is that it is difficult to find an ALP which is coupled to photons without also having a coupling to gluons, similar to the case in grand unified theories. This translates into a bound on the ALP parameter space:
\begin{align}
    \frac{\ga}{m_a} \leq 
    \frac{\alpha_\EM}{2\pi} \left( \frac{E}{N} -1.92 \right) 
    \frac{\left(\sqrt{z} + \frac{1}{\sqrt{z}}\right)}{f_\pi m_\pi}
    \, .
\end{align}

This has important implications for the current experimental program to search for axions, as most experiments rely on the axion coupling to photons. Many of the current and proposed experiments are sensitive to regions of parameter space above the QCD line which are not populated in theories where the axions couple equally to photons and gluons. If there was an experimental hint of a light axion detected through the photon coupling this would require a significant model building effort to explain in heterotic models. Such a signal would imply that the UV gauge group is $E_8 \times E_8$, the SM gauge groups contain factors from both $E_8$'s and the instanton contributions to the potential are sufficiently suppressed to lead to a light ALP. While these ingredients are in principle present in some orbifold compactifications, constructing realistic models where the SM is embedded in this way is a difficult task and to our knowledge there are no realistic models of this type. Heterotic models with the SM embedded into a single $E_8$ as well as $SO(32)$ and type I string theories would be incompatible with such a signal.

\section*{Acknowledgments}
We thank Kiwoon Choi, Naomi Gendler, Thomas Harvey, Andre Lukas, John March-Russell, Miguel Montero, Matthew McCullough, \textcolor{orange}{Arthur Platschorre}, Fernando Quevedo, Stefan Stelzl, and Irene Valenzuela for useful conversations and discussions. We also thank Luis Alvarez-Gaum\'e, Thomas Harvey, John March-Russell, Fernando Quevedo and Matt Reece for useful comments on a draft of the manuscript. PA is supported by the STFC under Grant No. ST/T000864/1. MN is supported in part by a joint Clarendon and Sloane-Robinson scholarship from Oxford University and Keble college, and by the National Science Foundation under Grant No. PHY-2310717. MR is supported by the STFC grant ST/T006242/1.

\appendix

\section{Contributions to axion potential}

\label{app:shifts}

In this section we summarise the possible contributions to the axion potential in addition to QCD instantons. These contributions are not necessarily aligned with the vacuum of the QCD-induced potential so can spoil the solution to the strong CP problem if they are not much smaller than QCD effects. This is the usual PQ quality problem, for which we require:
\begin{equation}
    V_{\cancel{PQ}}= -\Lambda_{\cancel{PQ}}^4\cos\left ( \frac{a}{f_a} +\delta \right )\,,\hspace{2cm}\frac{|V_{\cancel{PQ}}|}{m_\pi^2f_\pi^2}\lesssim 10^{-10}\,,
\end{equation}
where $a$ is the QCD axion. 

The first possible contribution to the potential is from confining gauge sectors. For instance, if the second $E_8$ is unbroken at the compactification scale it confines at some energy scale $\Lambda_c$ and will generate a potential for $\varphi_2$
\begin{align}
    V \sim \Lambda_c^4 \cos (\varphi_2) \, .
\end{align}
Even if the second $E_8$ is broken, there may be gauge sectors unbroken at the compactification scale which confine at low energies, generating a potential for either $\varphi_1$ or $\varphi_2$ depending on which $E_8$ factor they come from. If these confining sectors come from the first $E_8$ they could lead to a quality problem unless the new contributions to the potential have flat directions.

There are also UV instanton contributions to the potential. One example is euclidean NS5-branes wrapping the entire internal 6-dimensional space, $K$, which generate a potential for the model-independent axion. These kind of branes -- electrically charged under the 6-form $B_6$ -- are not described in 10D SUGRA but can be continuously deformed into Yang-Mills (YM) instantons at the compactification scale ($R^{-1}$)~\cite{Strominger:1990et}. The instanton action depends on the volume of the entire compact space:
\begin{equation}
    S_{NS5}\sim \frac{V_X}{g_s^2 l_s^6}\,.
\end{equation}
These instantons generate a potential for the model-independent axion which goes like
\begin{align}
    V \sim R^{-4} e^{-S_{NS5}} \cos\left( \frac{a}{f_a} +\delta_a \right)\, .
\end{align}
In the heterotic string theory the gauge coupling is given by:
\begin{equation}
    \frac{1}{\alpha_{YM}}=\frac{1}{g_s^2}\frac{V_X}{l_s^6}\simeq \frac{2\pi}{\alpha_{\text{GUT}}}\, ,
\end{equation}
so $S_{NS5}$ corresponds to the action of a YM instanton at the compactification scale $S_{YM}\sim \frac{2\pi}{\alpha_{YM}}$. 

On the other hand, there are also worldsheet instantons which generate a potential for the model-dependent axions~\cite{Wen:1985jz}. 
These correspond to euclidean closed string worldsheets wrapping 2-cycles. They are fully-localised solutions in Minkowski space and correspond to instanton-like effects breaking the (continuous) shift symmetries of the model dependent axions. Each model-dependent axion comes with an associated worldsheet instanton. Their effect is controlled by the action, $S_{ws}^{(i)}$, given by the volume of the wrapped 2-cycle: 
\begin{equation}
    S_{ws}^{(i)}=\frac{2\pi V_C^{(i)}}{l_s^2}\,.
\end{equation}
Similar to the NS5-instantons, these contributions are expected to be of the form
\begin{align}
    V \sim R^{-4} e^{-S_{ws}^{(i)}}  \cos\left( \frac{b_i}{f_{b_i}} +\delta_i \right)\, .
\end{align}
All in all, NS5 branes, gauge instantons and worldsheet instantons all contribute to the effective potential in equations~\eqref{eq:eff_Lag_E8_axion}~\&~\eqref{eq:eff_Lag_non-standard_embedding}, which generates mass terms for the axions.

\section{Twisted state contribution to the axion-photon coupling}
\label{app:twisted}

In a theory with discrete Wilson lines -- that is, on a non-simply connected space -- there can appear fractionally charged states that would not appear in a 4d GUT \cite{Wen:1985qj}. Their masses, roughly given by $\sim r/\alpha^\prime$, depend on the size of the non-contractible cycles. While they are generically heavy in CY manifolds, they can be light in orbifold compactifications, where the cycle shrinks to a singular point. One could therefore ask if, after integrating them out, these degrees of freedom induce a non-GUT universal axion-photon coupling, modifying the result Eq.\eqref{eq:eff_Lag_E8_axion}. In this section we compute the twisted state contribution to $\ga$ for CY and orbifold compactifications. If the twisted states are allowed to have a vector-like mass, their contribution is exponentially suppressed with respect to the leading contributions to $\ga\sim \frac{\alpha_\EM}{2\pi}\frac{1}{f_a}$. On the other hand, when PQ symmetry forbids a vector-like mass, their induced axion couplings are GUT-universal. 

The axions that we consider in this section are those coming from the $B$ field but similar arguments apply for field theory axions. In order to obtain an axion-photon coupling induced by a twisted state, we consider operators of the type:
\begin{equation}
    \mu\bar{\Psi} e^{ia\gamma_5}\Psi\,,
\end{equation}
where $\Psi$ is the twisted state, which we model by a charged, Dirac fermion of mass $m_\Psi$. 

Due to (higher-form) gauge invariance of the $B$ field any term involving the MI or MD axions in the 4d EFT comes with an exponential suppression $e^{-I}$.  The quantity $I$ is a volume factor which for the model-independent axion is the 6d volume, $V_{K}/l_s^6$, while for the case of model-dependent axions it is given by the volume of a 2-cycle, $V_C/l_s^2$. Another argument is that with $N=1$ SUSY, the superpotential is an holomorphic function of the complexified moduli, so any term involving $e^{ia}$ or $e^{ib}$ comes with an exponential $e^{-I}$ from the real part of the supermultiplet.

After integrating out $\Psi$, we obtain the coupling to photons~\cite{Agrawal:2023sbp}:
\begin{equation}
    \ga = \frac{\alpha_\EM}{2\pi f_a}z\,,\,\, \text{ with: }z=\frac{\mu}{m_\Psi}\,.
\end{equation}
As emphasised above, axion masses are protected  by the higher-form global symmetry of the $B$ field in 10d. This means that in the case of twisted states, the coupling is predicted to be exponentially suppressed:
\begin{equation}
    \mu \sim f_ae^{-I}\,.
\end{equation}

To perform an estimate of the induced $g_a/m_\Psi$ for a generic ALP, let us assume that it gets no additional potential from IR instantons (e.g. from QCD). UV instantons will always generate a potential of the type:
\begin{equation}
    V_{UV}(a)\sim R^{-4}e^{-I}\cos (a)\,.
\end{equation}
It is illustrative to take the following approximation: $R^{-1}\sim f_a$. In this case one can obtain a simple formula for the axion coupling induced by twisted states:
\begin{equation}
    \delta \ga\sim \frac{\alpha_\EM}{2\pi f_a}\frac{m_a}{m_\Psi}e^{-I/2}\,.
\end{equation}
One can easily see that the axion coupling is suppressed by the axion mass as well as by the exponentially small UV instanton effects.
This applies for both CY and orbifolds, the difference being the expected value  of the twisted state mass, $m_\Psi$. In CY one naturally expects $m_\Psi\sim R^{-1}$, while in orbifold compactifications twisted states may be light (they may correspond to SM particles). 
In either case, 
the axion-photon coupling generated by integrating out twisted states is exponentially suppressed with respect to the axion couplings obtained from the 10d Green-Schwarz mechanism.

\bibliographystyle{utphys}
\bibliography{refs}

\providecommand{\href}[2]{#2}\begingroup\raggedright\begin{thebibliography}{100}

\bibitem{Gross:1984dd}
D.~J. Gross, J.~A. Harvey, E.~J. Martinec, and R.~Rohm, ``{The Heterotic
  String},'' \href{http://dx.doi.org/10.1103/PhysRevLett.54.502}{{\em Phys.
  Rev. Lett.} {\bfseries 54} (1985) 502--505}.

\bibitem{Green:1984sg}
M.~B. Green and J.~H. Schwarz, ``{Anomaly Cancellation in Supersymmetric D=10
  Gauge Theory and Superstring Theory},''
  \href{http://dx.doi.org/10.1016/0370-2693(84)91565-X}{{\em Phys. Lett. B}
  {\bfseries 149} (1984) 117--122}.

\bibitem{Alvarez-Gaume:1986ghj}
L.~Alvarez-Gaume, P.~H. Ginsparg, G.~W. Moore, and C.~Vafa, ``{An O(16) x O(16)
  Heterotic String},''
  \href{http://dx.doi.org/10.1016/0370-2693(86)91524-8}{{\em Phys. Lett. B}
  {\bfseries 171} (1986) 155--162}.

\bibitem{Witten:1995gx}
E.~Witten, ``{Small instantons in string theory},''
  \href{http://dx.doi.org/10.1016/0550-3213(95)00625-7}{{\em Nucl. Phys. B}
  {\bfseries 460} (1996) 541--559},
  \href{http://arxiv.org/abs/hep-th/9511030}{{\ttfamily arXiv:hep-th/9511030}}.

\bibitem{Benakli:1999yc}
K.~Benakli and Y.~Oz, ``{Small instantons and weak scale string theory},''
  \href{http://dx.doi.org/10.1016/S0370-2693(99)01422-7}{{\em Phys. Lett. B}
  {\bfseries 472} (2000) 83--88},
  \href{http://arxiv.org/abs/hep-th/9910090}{{\ttfamily arXiv:hep-th/9910090}}.

\bibitem{Candelas:1985en}
P.~Candelas, G.~T. Horowitz, A.~Strominger, and E.~Witten, ``{Vacuum
  configurations for superstrings},''
  \href{http://dx.doi.org/10.1016/0550-3213(85)90602-9}{{\em Nucl. Phys. B}
  {\bfseries 258} (1985) 46--74}.

\bibitem{Dixon:1985jw}
L.~J. Dixon, J.~A. Harvey, C.~Vafa, and E.~Witten, ``{Strings on Orbifolds},''
  \href{http://dx.doi.org/10.1016/0550-3213(85)90593-0}{{\em Nucl. Phys. B}
  {\bfseries 261} (1985) 678--686}.

\bibitem{Dixon:1986jc}
L.~J. Dixon, J.~A. Harvey, C.~Vafa, and E.~Witten, ``{Strings on Orbifolds.
  2.},'' \href{http://dx.doi.org/10.1016/0550-3213(86)90287-7}{{\em Nucl. Phys.
  B} {\bfseries 274} (1986) 285--314}.

\bibitem{Ibanez:1986tp}
L.~E. Ibanez, H.~P. Nilles, and F.~Quevedo, ``{Orbifolds and Wilson Lines},''
  \href{http://dx.doi.org/10.1016/0370-2693(87)90066-9}{{\em Phys. Lett. B}
  {\bfseries 187} (1987) 25--32}.

\bibitem{Ibanez:1987xa}
L.~E. Ibanez, H.~P. Nilles, and F.~Quevedo, ``{Reducing the Rank of the Gauge
  Group in Orbifold Compactifications of the Heterotic String},''
  \href{http://dx.doi.org/10.1016/0370-2693(87)90117-1}{{\em Phys. Lett. B}
  {\bfseries 192} (1987) 332--338}.

\bibitem{Witten:1985xc}
E.~Witten, ``{Symmetry breaking patterns in superstring models},''
  \href{http://dx.doi.org/10.1016/0550-3213(85)90603-0}{{\em Nucl. Phys. B}
  {\bfseries 258} (1985) 75--100}.

\bibitem{Svrcek:2006yi}
P.~Svrcek and E.~Witten, ``{Axions In String Theory},''
  \href{http://dx.doi.org/10.1088/1126-6708/2006/06/051}{{\em JHEP} {\bfseries
  06} (2006) 051}, \href{http://arxiv.org/abs/hep-th/0605206}{{\ttfamily
  arXiv:hep-th/0605206}}.

\bibitem{Arvanitaki:2009fg}
A.~Arvanitaki, S.~Dimopoulos, S.~Dubovsky, N.~Kaloper, and J.~March-Russell,
  ``{String Axiverse},''
  \href{http://dx.doi.org/10.1103/PhysRevD.81.123530}{{\em Phys. Rev. D}
  {\bfseries 81} (2010) 123530},
  \href{http://arxiv.org/abs/0905.4720}{{\ttfamily arXiv:0905.4720 [hep-th]}}.

\bibitem{Kamionkowski:1992mf}
M.~Kamionkowski and J.~March-Russell, ``{Planck scale physics and the
  Peccei-Quinn mechanism},''
  \href{http://dx.doi.org/10.1016/0370-2693(92)90492-M}{{\em Phys. Lett. B}
  {\bfseries 282} (1992) 137--141},
  \href{http://arxiv.org/abs/hep-th/9202003}{{\ttfamily arXiv:hep-th/9202003}}.

\bibitem{Beyer:2022ywc}
K.~A. Beyer and S.~Sarkar, ``{Ruling out light axions: The writing is on the
  wall},'' \href{http://dx.doi.org/10.21468/SciPostPhys.15.1.003}{{\em SciPost
  Phys.} {\bfseries 15} no.~1, (2023) 003},
  \href{http://arxiv.org/abs/2211.14635}{{\ttfamily arXiv:2211.14635
  [hep-ph]}}.

\bibitem{Lu:2023ayc}
Q.~Lu, M.~Reece, and Z.~Sun, ``{The quality/cosmology tension for a
  post-inflation QCD axion},''
  \href{http://dx.doi.org/10.1007/JHEP07(2024)227}{{\em JHEP} {\bfseries 07}
  (2024) 227}, \href{http://arxiv.org/abs/2312.07650}{{\ttfamily
  arXiv:2312.07650 [hep-ph]}}.

\bibitem{Craig:2024dnl}
N.~Craig and M.~Kongsore, ``{High-Quality Axions from Higher-Form Symmetries in
  Extra Dimensions},'' \href{http://arxiv.org/abs/2408.10295}{{\ttfamily
  arXiv:2408.10295 [hep-ph]}}.

\bibitem{Agrawal:2017cmd}
P.~Agrawal, J.~Fan, M.~Reece, and L.-T. Wang, ``{Experimental Targets for
  Photon Couplings of the QCD Axion},''
  \href{http://dx.doi.org/10.1007/JHEP02(2018)006}{{\em JHEP} {\bfseries 02}
  (2018) 006}, \href{http://arxiv.org/abs/1709.06085}{{\ttfamily
  arXiv:1709.06085 [hep-ph]}}.

\bibitem{Fraser:2019ojt}
K.~Fraser and M.~Reece, ``{Axion Periodicity and Coupling Quantization in the
  Presence of Mixing},'' \href{http://dx.doi.org/10.1007/JHEP05(2020)066}{{\em
  JHEP} {\bfseries 05} (2020) 066},
  \href{http://arxiv.org/abs/1910.11349}{{\ttfamily arXiv:1910.11349
  [hep-ph]}}.

\bibitem{Reece:2023iqn}
M.~Reece, ``{Axion-gauge coupling quantization with a twist},''
  \href{http://dx.doi.org/10.1007/JHEP10(2023)116}{{\em JHEP} {\bfseries 10}
  (2023) 116}, \href{http://arxiv.org/abs/2309.03939}{{\ttfamily
  arXiv:2309.03939 [hep-ph]}}.

\bibitem{Cordova:2023her}
C.~Cordova, S.~Hong, and L.-T. Wang, ``{Axion domain walls, small instantons,
  and non-invertible symmetry breaking},''
  \href{http://dx.doi.org/10.1007/JHEP05(2024)325}{{\em JHEP} {\bfseries 05}
  (2024) 325}, \href{http://arxiv.org/abs/2309.05636}{{\ttfamily
  arXiv:2309.05636 [hep-ph]}}.

\bibitem{Agrawal:2023sbp}
P.~Agrawal and A.~Platschorre, ``{The monodromic axion-photon coupling},''
  \href{http://dx.doi.org/10.1007/JHEP01(2024)169}{{\em JHEP} {\bfseries 01}
  (2024) 169}, \href{http://arxiv.org/abs/2309.03934}{{\ttfamily
  arXiv:2309.03934 [hep-th]}}.

\bibitem{Choi:2023pdp}
Y.~Choi, M.~Forslund, H.~T. Lam, and S.-H. Shao, ``{Quantization of Axion-Gauge
  Couplings and Noninvertible Higher Symmetries},''
  \href{http://dx.doi.org/10.1103/PhysRevLett.132.121601}{{\em Phys. Rev.
  Lett.} {\bfseries 132} no.~12, (2024) 121601},
  \href{http://arxiv.org/abs/2309.03937}{{\ttfamily arXiv:2309.03937
  [hep-ph]}}.

\bibitem{Agrawal:2019lkr}
P.~Agrawal, A.~Hook, and J.~Huang, ``{A CMB Millikan experiment with cosmic
  axiverse strings},'' \href{http://dx.doi.org/10.1007/JHEP07(2020)138}{{\em
  JHEP} {\bfseries 07} (2020) 138},
  \href{http://arxiv.org/abs/1912.02823}{{\ttfamily arXiv:1912.02823
  [astro-ph.CO]}}.

\bibitem{Yin:2023vit}
W.~W. Yin, L.~Dai, and S.~Ferraro, ``{Testing charge quantization with axion
  string-induced cosmic birefringence},''
  \href{http://dx.doi.org/10.1088/1475-7516/2023/07/052}{{\em JCAP} {\bfseries
  07} (2023) 052}, \href{http://arxiv.org/abs/2305.02318}{{\ttfamily
  arXiv:2305.02318 [astro-ph.CO]}}.

\bibitem{Agrawal:2022lsp}
P.~Agrawal, M.~Nee, and M.~Reig, ``{Axion couplings in grand unified
  theories},'' \href{http://dx.doi.org/10.1007/JHEP10(2022)141}{{\em JHEP}
  {\bfseries 10} (2022) 141}, \href{http://arxiv.org/abs/2206.07053}{{\ttfamily
  arXiv:2206.07053 [hep-ph]}}.

\bibitem{Ibanez:1987sn}
L.~E. Ibanez, J.~E. Kim, H.~P. Nilles, and F.~Quevedo, ``{Orbifold
  Compactifications with Three Families of SU(3) x SU(2) x U(1)**n},''
  \href{http://dx.doi.org/10.1016/0370-2693(87)90255-3}{{\em Phys. Lett. B}
  {\bfseries 191} (1987) 282--286}.

\bibitem{Font:1989aj}
A.~Font, L.~E. Ibanez, F.~Quevedo, and A.~Sierra, ``{The Construction of
  'Realistic' Four-Dimensional Strings Through Orbifolds},''
  \href{http://dx.doi.org/10.1016/0550-3213(90)90215-Y}{{\em Nucl. Phys. B}
  {\bfseries 331} (1990) 421--474}.

\bibitem{Choi:2006qj}
K.-S. Choi, I.-W. Kim, and J.~E. Kim, ``{String compactification, QCD axion and
  axion-photon-photon coupling},''
  \href{http://dx.doi.org/10.1088/1126-6708/2007/03/116}{{\em JHEP} {\bfseries
  03} (2007) 116}, \href{http://arxiv.org/abs/hep-ph/0612107}{{\ttfamily
  arXiv:hep-ph/0612107}}.

\bibitem{Choi:2009jt}
K.-S. Choi, H.~P. Nilles, S.~Ramos-Sanchez, and P.~K.~S. Vaudrevange,
  ``{Accions},'' \href{http://dx.doi.org/10.1016/j.physletb.2009.04.028}{{\em
  Phys. Lett. B} {\bfseries 675} (2009) 381--386},
  \href{http://arxiv.org/abs/0902.3070}{{\ttfamily arXiv:0902.3070 [hep-th]}}.

\bibitem{Blumenhagen:2005ga}
R.~Blumenhagen, G.~Honecker, and T.~Weigand, ``{Loop-corrected
  compactifications of the heterotic string with line bundles},''
  \href{http://dx.doi.org/10.1088/1126-6708/2005/06/020}{{\em JHEP} {\bfseries
  06} (2005) 020}, \href{http://arxiv.org/abs/hep-th/0504232}{{\ttfamily
  arXiv:hep-th/0504232}}.

\bibitem{Anderson:2011ns}
L.~B. Anderson, J.~Gray, A.~Lukas, and E.~Palti, ``{Two Hundred Heterotic
  Standard Models on Smooth Calabi-Yau Threefolds},''
  \href{http://dx.doi.org/10.1103/PhysRevD.84.106005}{{\em Phys. Rev. D}
  {\bfseries 84} (2011) 106005},
  \href{http://arxiv.org/abs/1106.4804}{{\ttfamily arXiv:1106.4804 [hep-th]}}.

\bibitem{Sikivie:1983ip}
P.~Sikivie, ``{Experimental Tests of the Invisible Axion},''
  \href{http://dx.doi.org/10.1103/PhysRevLett.51.1415}{{\em Phys. Rev. Lett.}
  {\bfseries 51} (1983) 1415--1417}. [Erratum: Phys.Rev.Lett. 52, 695 (1984)].

\bibitem{Sikivie:1985yu}
P.~Sikivie, ``{Detection Rates for 'Invisible' Axion Searches},''
  \href{http://dx.doi.org/10.1103/PhysRevD.36.974}{{\em Phys. Rev. D}
  {\bfseries 32} (1985) 2988}. [Erratum: Phys.Rev.D 36, 974 (1987)].

\bibitem{Wilczek:1987mv}
F.~Wilczek, ``{Two Applications of Axion Electrodynamics},''
  \href{http://dx.doi.org/10.1103/PhysRevLett.58.1799}{{\em Phys. Rev. Lett.}
  {\bfseries 58} (1987) 1799}.

\bibitem{CAST:2004gzq}
{\bfseries CAST} Collaboration, K.~Zioutas {\em et~al.}, ``{First results from
  the CERN Axion Solar Telescope (CAST)},''
  \href{http://dx.doi.org/10.1103/PhysRevLett.94.121301}{{\em Phys. Rev. Lett.}
  {\bfseries 94} (2005) 121301},
  \href{http://arxiv.org/abs/hep-ex/0411033}{{\ttfamily arXiv:hep-ex/0411033}}.

\bibitem{Kahn:2016aff}
Y.~Kahn, B.~R. Safdi, and J.~Thaler, ``{Broadband and Resonant Approaches to
  Axion Dark Matter Detection},''
  \href{http://dx.doi.org/10.1103/PhysRevLett.117.141801}{{\em Phys. Rev.
  Lett.} {\bfseries 117} no.~14, (2016) 141801},
  \href{http://arxiv.org/abs/1602.01086}{{\ttfamily arXiv:1602.01086
  [hep-ph]}}.

\bibitem{Arvanitaki:2017nhi}
A.~Arvanitaki, S.~Dimopoulos, and K.~Van~Tilburg, ``{Resonant absorption of
  bosonic dark matter in molecules},''
  \href{http://dx.doi.org/10.1103/PhysRevX.8.041001}{{\em Phys. Rev. X}
  {\bfseries 8} no.~4, (2018) 041001},
  \href{http://arxiv.org/abs/1709.05354}{{\ttfamily arXiv:1709.05354
  [hep-ph]}}.

\bibitem{Baryakhtar:2018doz}
M.~Baryakhtar, J.~Huang, and R.~Lasenby, ``{Axion and hidden photon dark matter
  detection with multilayer optical haloscopes},''
  \href{http://dx.doi.org/10.1103/PhysRevD.98.035006}{{\em Phys. Rev. D}
  {\bfseries 98} no.~3, (2018) 035006},
  \href{http://arxiv.org/abs/1803.11455}{{\ttfamily arXiv:1803.11455
  [hep-ph]}}.

\bibitem{Chaudhuri:2018rqn}
S.~Chaudhuri, K.~Irwin, P.~W. Graham, and J.~Mardon, ``{Optimal Impedance
  Matching and Quantum Limits of Electromagnetic Axion and Hidden-Photon Dark
  Matter Searches},'' \href{http://arxiv.org/abs/1803.01627}{{\ttfamily
  arXiv:1803.01627 [hep-ph]}}.

\bibitem{Fedderke:2019ajk}
M.~A. Fedderke, P.~W. Graham, and S.~Rajendran, ``{Axion Dark Matter Detection
  with CMB Polarization},''
  \href{http://dx.doi.org/10.1103/PhysRevD.100.015040}{{\em Phys. Rev. D}
  {\bfseries 100} no.~1, (2019) 015040},
  \href{http://arxiv.org/abs/1903.02666}{{\ttfamily arXiv:1903.02666
  [astro-ph.CO]}}.

\bibitem{Langhoff:2022bij}
K.~Langhoff, N.~J. Outmezguine, and N.~L. Rodd, ``{Irreducible Axion
  Background},'' \href{http://dx.doi.org/10.1103/PhysRevLett.129.241101}{{\em
  Phys. Rev. Lett.} {\bfseries 129} no.~24, (2022) 241101},
  \href{http://arxiv.org/abs/2209.06216}{{\ttfamily arXiv:2209.06216
  [hep-ph]}}.

\bibitem{QSHS:2023jny}
{\bfseries QSHS} Collaboration, I.~Bailey {\em et~al.}, ``{Searching for
  wave-like dark matter with QSHS},''
  \href{http://dx.doi.org/10.21468/SciPostPhysProc.12.040}{{\em SciPost Phys.
  Proc.} {\bfseries 12} (2023) 040}.

\bibitem{Moody:1984ba}
J.~E. Moody and F.~Wilczek, ``{NEW MACROSCOPIC FORCES?},''
  \href{http://dx.doi.org/10.1103/PhysRevD.30.130}{{\em Phys. Rev. D}
  {\bfseries 30} (1984) 130}.

\bibitem{Graham:2011qk}
P.~W. Graham and S.~Rajendran, ``{Axion Dark Matter Detection with Cold
  Molecules},'' \href{http://dx.doi.org/10.1103/PhysRevD.84.055013}{{\em Phys.
  Rev. D} {\bfseries 84} (2011) 055013},
  \href{http://arxiv.org/abs/1101.2691}{{\ttfamily arXiv:1101.2691 [hep-ph]}}.

\bibitem{Budker:2013hfa}
D.~Budker, P.~W. Graham, M.~Ledbetter, S.~Rajendran, and A.~Sushkov,
  ``{Proposal for a Cosmic Axion Spin Precession Experiment (CASPEr)},''
  \href{http://dx.doi.org/10.1103/PhysRevX.4.021030}{{\em Phys. Rev. X}
  {\bfseries 4} no.~2, (2014) 021030},
  \href{http://arxiv.org/abs/1306.6089}{{\ttfamily arXiv:1306.6089 [hep-ph]}}.

\bibitem{Graham:2013gfa}
P.~W. Graham and S.~Rajendran, ``{New Observables for Direct Detection of Axion
  Dark Matter},'' \href{http://dx.doi.org/10.1103/PhysRevD.88.035023}{{\em
  Phys. Rev. D} {\bfseries 88} (2013) 035023},
  \href{http://arxiv.org/abs/1306.6088}{{\ttfamily arXiv:1306.6088 [hep-ph]}}.

\bibitem{Armengaud:2013rta}
E.~Armengaud {\em et~al.}, ``{Axion searches with the EDELWEISS-II
  experiment},'' \href{http://dx.doi.org/10.1088/1475-7516/2013/11/067}{{\em
  JCAP} {\bfseries 11} (2013) 067},
  \href{http://arxiv.org/abs/1307.1488}{{\ttfamily arXiv:1307.1488
  [astro-ph.CO]}}.

\bibitem{Arvanitaki:2014dfa}
A.~Arvanitaki and A.~A. Geraci, ``{Resonantly Detecting Axion-Mediated Forces
  with Nuclear Magnetic Resonance},''
  \href{http://dx.doi.org/10.1103/PhysRevLett.113.161801}{{\em Phys. Rev.
  Lett.} {\bfseries 113} no.~16, (2014) 161801},
  \href{http://arxiv.org/abs/1403.1290}{{\ttfamily arXiv:1403.1290 [hep-ph]}}.

\bibitem{Arvanitaki:2021wjk}
A.~Arvanitaki, A.~Madden, and K.~Van~Tilburg, ``{Piezoaxionic effect},''
  \href{http://dx.doi.org/10.1103/PhysRevD.109.072009}{{\em Phys. Rev. D}
  {\bfseries 109} no.~7, (2024) 072009},
  \href{http://arxiv.org/abs/2112.11466}{{\ttfamily arXiv:2112.11466
  [hep-ph]}}.

\bibitem{Berlin:2022mia}
A.~Berlin and K.~Zhou, ``{Discovering QCD-coupled axion dark matter with
  polarization haloscopes},''
  \href{http://dx.doi.org/10.1103/PhysRevD.108.035038}{{\em Phys. Rev. D}
  {\bfseries 108} no.~3, (2023) 035038},
  \href{http://arxiv.org/abs/2209.12901}{{\ttfamily arXiv:2209.12901
  [hep-ph]}}.

\bibitem{Graham:2015ouw}
P.~W. Graham, I.~G. Irastorza, S.~K. Lamoreaux, A.~Lindner, and K.~A. van
  Bibber, ``{Experimental Searches for the Axion and Axion-Like Particles},''
  \href{http://dx.doi.org/10.1146/annurev-nucl-102014-022120}{{\em Ann. Rev.
  Nucl. Part. Sci.} {\bfseries 65} (2015) 485--514},
  \href{http://arxiv.org/abs/1602.00039}{{\ttfamily arXiv:1602.00039
  [hep-ex]}}.

\bibitem{Irastorza:2018dyq}
I.~G. Irastorza and J.~Redondo, ``{New experimental approaches in the search
  for axion-like particles},''
  \href{http://dx.doi.org/10.1016/j.ppnp.2018.05.003}{{\em Prog. Part. Nucl.
  Phys.} {\bfseries 102} (2018) 89--159},
  \href{http://arxiv.org/abs/1801.08127}{{\ttfamily arXiv:1801.08127
  [hep-ph]}}.

\bibitem{OHare:2024nmr}
C.~A.~J. O'Hare, ``{Cosmology of axion dark matter},''
  \href{http://dx.doi.org/10.22323/1.454.0040}{{\em PoS} {\bfseries
  COSMICWISPers} (2024) 040}, \href{http://arxiv.org/abs/2403.17697}{{\ttfamily
  arXiv:2403.17697 [hep-ph]}}.

\bibitem{Minami:2020odp}
Y.~Minami and E.~Komatsu, ``{New Extraction of the Cosmic Birefringence from
  the Planck 2018 Polarization Data},''
  \href{http://dx.doi.org/10.1103/PhysRevLett.125.221301}{{\em Phys. Rev.
  Lett.} {\bfseries 125} no.~22, (2020) 221301},
  \href{http://arxiv.org/abs/2011.11254}{{\ttfamily arXiv:2011.11254
  [astro-ph.CO]}}.

\bibitem{BICEPKeck:2021sbt}
{\bfseries BICEP/Keck} Collaboration, P.~A.~R. Ade {\em et~al.}, ``{BICEP/Keck
  XIV: Improved constraints on axionlike polarization oscillations in the
  cosmic microwave background},''
  \href{http://dx.doi.org/10.1103/PhysRevD.105.022006}{{\em Phys. Rev. D}
  {\bfseries 105} no.~2, (2022) 022006},
  \href{http://arxiv.org/abs/2108.03316}{{\ttfamily arXiv:2108.03316
  [astro-ph.CO]}}.

\bibitem{SPT-3G:2022ods}
{\bfseries SPT-3G} Collaboration, K.~R. Ferguson {\em et~al.}, ``{Searching for
  axionlike time-dependent cosmic birefringence with data from SPT-3G},''
  \href{http://dx.doi.org/10.1103/PhysRevD.106.042011}{{\em Phys. Rev. D}
  {\bfseries 106} no.~4, (2022) 042011},
  \href{http://arxiv.org/abs/2203.16567}{{\ttfamily arXiv:2203.16567
  [astro-ph.CO]}}.

\bibitem{Eskilt:2022cff}
J.~R. Eskilt and E.~Komatsu, ``{Improved constraints on cosmic birefringence
  from the WMAP and Planck cosmic microwave background polarization data},''
  \href{http://dx.doi.org/10.1103/PhysRevD.106.063503}{{\em Phys. Rev. D}
  {\bfseries 106} no.~6, (2022) 063503},
  \href{http://arxiv.org/abs/2205.13962}{{\ttfamily arXiv:2205.13962
  [astro-ph.CO]}}.

\bibitem{Nakai:2023zdr}
Y.~Nakai, R.~Namba, I.~Obata, Y.-C. Qiu, and R.~Saito, ``{Can we explain cosmic
  birefringence without a new light field beyond Standard Model?},''
  \href{http://dx.doi.org/10.1007/JHEP01(2024)057}{{\em JHEP} {\bfseries 01}
  (2024) 057}, \href{http://arxiv.org/abs/2310.09152}{{\ttfamily
  arXiv:2310.09152 [astro-ph.CO]}}.

\bibitem{Diego-Palazuelos:2023mpy}
P.~Diego-Palazuelos, ``{Search for ultra-light axions with CMB polarization},''
\newblock 4, 2023.
\newblock \href{http://arxiv.org/abs/2304.03647}{{\ttfamily arXiv:2304.03647
  [astro-ph.CO]}}.

\bibitem{Cosmoglobe:2023pgf}
{\bfseries Cosmoglobe} Collaboration, J.~R. Eskilt {\em et~al.}, ``{COSMOGLOBE
  DR1 results - II. Constraints on isotropic cosmic birefringence from
  reprocessed WMAP and Planck LFI data},''
  \href{http://dx.doi.org/10.1051/0004-6361/202346829}{{\em Astron. Astrophys.}
  {\bfseries 679} (2023) A144},
  \href{http://arxiv.org/abs/2305.02268}{{\ttfamily arXiv:2305.02268
  [astro-ph.CO]}}.

\bibitem{POLARBEAR:2023ric}
{\bfseries POLARBEAR} Collaboration, S.~Adachi {\em et~al.}, ``{Constraints on
  axionlike polarization oscillations in the cosmic microwave background with
  POLARBEAR},'' \href{http://dx.doi.org/10.1103/PhysRevD.108.043017}{{\em Phys.
  Rev. D} {\bfseries 108} no.~4, (2023) 043017},
  \href{http://arxiv.org/abs/2303.08410}{{\ttfamily arXiv:2303.08410
  [astro-ph.CO]}}.

\bibitem{POLARBEAR:2024vel}
{\bfseries POLARBEAR} Collaboration, S.~Adachi {\em et~al.}, ``{Exploration of
  the polarization angle variability of the Crab Nebula with POLARBEAR and its
  application to the search for axionlike particles},''
  \href{http://dx.doi.org/10.1103/PhysRevD.110.063013}{{\em Phys. Rev. D}
  {\bfseries 110} no.~6, (2024) 063013},
  \href{http://arxiv.org/abs/2403.02096}{{\ttfamily arXiv:2403.02096
  [astro-ph.CO]}}.

\bibitem{Arvanitaki:2010sy}
A.~Arvanitaki and S.~Dubovsky, ``{Exploring the String Axiverse with Precision
  Black Hole Physics},''
  \href{http://dx.doi.org/10.1103/PhysRevD.83.044026}{{\em Phys. Rev. D}
  {\bfseries 83} (2011) 044026},
  \href{http://arxiv.org/abs/1004.3558}{{\ttfamily arXiv:1004.3558 [hep-th]}}.

\bibitem{Wouters:2013hua}
D.~Wouters and P.~Brun, ``{Constraints on Axion-like Particles from X-Ray
  Observations of the Hydra Galaxy Cluster},''
  \href{http://dx.doi.org/10.1088/0004-637X/772/1/44}{{\em Astrophys. J.}
  {\bfseries 772} (2013) 44}, \href{http://arxiv.org/abs/1304.0989}{{\ttfamily
  arXiv:1304.0989 [astro-ph.HE]}}.

\bibitem{Marsh:2017yvc}
M.~C.~D. Marsh, H.~R. Russell, A.~C. Fabian, B.~P. McNamara, P.~Nulsen, and
  C.~S. Reynolds, ``{A New Bound on Axion-Like Particles},''
  \href{http://dx.doi.org/10.1088/1475-7516/2017/12/036}{{\em JCAP} {\bfseries
  12} (2017) 036}, \href{http://arxiv.org/abs/1703.07354}{{\ttfamily
  arXiv:1703.07354 [hep-ph]}}.

\bibitem{Reynolds:2019uqt}
C.~S. Reynolds, M.~C.~D. Marsh, H.~R. Russell, A.~C. Fabian, R.~Smith,
  F.~Tombesi, and S.~Veilleux, ``{Astrophysical limits on very light axion-like
  particles from Chandra grating spectroscopy of NGC 1275},''
  \href{http://dx.doi.org/10.3847/1538-4357/ab6a0c}{{\em Astrophys. J.}
  {\bfseries 890} (2020) 59}, \href{http://arxiv.org/abs/1907.05475}{{\ttfamily
  arXiv:1907.05475 [hep-ph]}}.

\bibitem{Reynes:2021bpe}
J.~S. Reyn\'es, J.~H. Matthews, C.~S. Reynolds, H.~R. Russell, R.~N. Smith, and
  M.~C.~D. Marsh, ``{New constraints on light axion-like particles using
  Chandra transmission grating spectroscopy of the powerful cluster-hosted
  quasar H1821+643},'' \href{http://dx.doi.org/10.1093/mnras/stab3464}{{\em
  Mon. Not. Roy. Astron. Soc.} {\bfseries 510} no.~1, (2021) 1264--1277},
  \href{http://arxiv.org/abs/2109.03261}{{\ttfamily arXiv:2109.03261
  [astro-ph.HE]}}.

\bibitem{Marsh:2018dlj}
D.~J.~E. Marsh, K.-C. Fong, E.~W. Lentz, L.~Smejkal, and M.~N. Ali, ``{Proposal
  to Detect Dark Matter using Axionic Topological Antiferromagnets},''
  \href{http://dx.doi.org/10.1103/PhysRevLett.123.121601}{{\em Phys. Rev.
  Lett.} {\bfseries 123} no.~12, (2019) 121601},
  \href{http://arxiv.org/abs/1807.08810}{{\ttfamily arXiv:1807.08810
  [hep-ph]}}.

\bibitem{Lawson:2019brd}
M.~Lawson, A.~J. Millar, M.~Pancaldi, E.~Vitagliano, and F.~Wilczek, ``{Tunable
  axion plasma haloscopes},''
  \href{http://dx.doi.org/10.1103/PhysRevLett.123.141802}{{\em Phys. Rev.
  Lett.} {\bfseries 123} no.~14, (2019) 141802},
  \href{http://arxiv.org/abs/1904.11872}{{\ttfamily arXiv:1904.11872
  [hep-ph]}}.

\bibitem{Beurthey:2020yuq}
S.~Beurthey {\em et~al.}, ``{MADMAX Status Report},''
  \href{http://arxiv.org/abs/2003.10894}{{\ttfamily arXiv:2003.10894
  [physics.ins-det]}}.

\bibitem{Schutte-Engel:2021bqm}
J.~Sch\"utte-Engel, D.~J.~E. Marsh, A.~J. Millar, A.~Sekine, F.~Chadha-Day,
  S.~Hoof, M.~N. Ali, K.-C. Fong, E.~Hardy, and L.~\v{S}mejkal, ``{Axion
  quasiparticles for axion dark matter detection},''
  \href{http://dx.doi.org/10.1088/1475-7516/2021/08/066}{{\em JCAP} {\bfseries
  08} (2021) 066}, \href{http://arxiv.org/abs/2102.05366}{{\ttfamily
  arXiv:2102.05366 [hep-ph]}}.

\bibitem{DMRadio:2022pkf}
{\bfseries DMRadio} Collaboration, L.~Brouwer {\em et~al.}, ``{Projected
  sensitivity of DMRadio-m3: A search for the QCD axion below
  1\,\,\ensuremath{\mu}eV},''
  \href{http://dx.doi.org/10.1103/PhysRevD.106.103008}{{\em Phys. Rev. D}
  {\bfseries 106} no.~10, (2022) 103008},
  \href{http://arxiv.org/abs/2204.13781}{{\ttfamily arXiv:2204.13781
  [hep-ex]}}.

\bibitem{Aja:2022csb}
B.~Aja {\em et~al.}, ``{The Canfranc Axion Detection Experiment (CADEx): search
  for axions at 90 GHz with Kinetic Inductance Detectors},''
  \href{http://dx.doi.org/10.1088/1475-7516/2022/11/044}{{\em JCAP} {\bfseries
  11} (2022) 044}, \href{http://arxiv.org/abs/2206.02980}{{\ttfamily
  arXiv:2206.02980 [hep-ex]}}.

\bibitem{Bourhill:2022alm}
J.~F. Bourhill, E.~C.~I. Paterson, M.~Goryachev, and M.~E. Tobar, ``{Searching
  for ultralight axions with twisted cavity resonators of anyon rotational
  symmetry with bulk modes of nonzero helicity},''
  \href{http://dx.doi.org/10.1103/PhysRevD.108.052014}{{\em Phys. Rev. D}
  {\bfseries 108} no.~5, (2023) 052014},
  \href{http://arxiv.org/abs/2208.01640}{{\ttfamily arXiv:2208.01640
  [hep-ph]}}.

\bibitem{ALPHA:2022rxj}
{\bfseries ALPHA} Collaboration, A.~J. Millar {\em et~al.}, ``{Searching for
  dark matter with plasma haloscopes},''
  \href{http://dx.doi.org/10.1103/PhysRevD.107.055013}{{\em Phys. Rev. D}
  {\bfseries 107} no.~5, (2023) 055013},
  \href{http://arxiv.org/abs/2210.00017}{{\ttfamily arXiv:2210.00017
  [hep-ph]}}.

\bibitem{ADMX:2023rsk}
{\bfseries ADMX} Collaboration, T.~Nitta {\em et~al.}, ``{Search for a
  Dark-Matter-Induced Cosmic Axion Background with ADMX},''
  \href{http://dx.doi.org/10.1103/PhysRevLett.131.101002}{{\em Phys. Rev.
  Lett.} {\bfseries 131} no.~10, (2023) 101002},
  \href{http://arxiv.org/abs/2303.06282}{{\ttfamily arXiv:2303.06282
  [hep-ex]}}.

\bibitem{Oshima:2023csb}
Y.~Oshima, H.~Fujimoto, J.~Kume, S.~Morisaki, K.~Nagano, T.~Fujita, I.~Obata,
  A.~Nishizawa, Y.~Michimura, and M.~Ando, ``{First results of axion dark
  matter search with DANCE},''
  \href{http://dx.doi.org/10.1103/PhysRevD.108.072005}{{\em Phys. Rev. D}
  {\bfseries 108} no.~7, (2023) 072005},
  \href{http://arxiv.org/abs/2303.03594}{{\ttfamily arXiv:2303.03594
  [hep-ex]}}.

\bibitem{DeMiguel:2023nmz}
{\bfseries DALI} Collaboration, J.~De~Miguel, J.~F. Hern\'andez-Cabrera,
  E.~Hern\'andez-Su\'arez, E.~Joven-\'Alvarez, C.~Otani, and J.~A. Rubi\~no
  Mart\'\i{}n, ``{Discovery prospects with the Dark-photons \& Axion-like
  particles Interferometer},''
  \href{http://dx.doi.org/10.1103/PhysRevD.109.062002}{{\em Phys. Rev. D}
  {\bfseries 109} no.~6, (2024) 062002},
  \href{http://arxiv.org/abs/2303.03997}{{\ttfamily arXiv:2303.03997
  [hep-ph]}}.

\bibitem{Ahyoune:2023gfw}
S.~Ahyoune {\em et~al.}, ``{A Proposal for a Low-Frequency Axion Search in the
  1\textendash{}2 \ensuremath{\mu}$\mu$ eV Range and Below with the BabyIAXO
  Magnet},'' \href{http://dx.doi.org/10.1002/andp.202300326}{{\em Annalen
  Phys.} {\bfseries 535} no.~12, (2023) 2300326},
  \href{http://arxiv.org/abs/2306.17243}{{\ttfamily arXiv:2306.17243
  [physics.ins-det]}}.

\bibitem{Alesini:2023qed}
D.~Alesini {\em et~al.}, ``{The future search for low-frequency axions and new
  physics with the FLASH resonant cavity experiment at Frascati National
  Laboratories},'' \href{http://dx.doi.org/10.1016/j.dark.2023.101370}{{\em
  Phys. Dark Univ.} {\bfseries 42} (2023) 101370},
  \href{http://arxiv.org/abs/2309.00351}{{\ttfamily arXiv:2309.00351
  [physics.ins-det]}}.

\bibitem{BREAD:2023xhc}
{\bfseries BREAD} Collaboration, S.~Knirck {\em et~al.}, ``{First Results from
  a Broadband Search for Dark Photon Dark Matter in the 44 to
  52\,\,\ensuremath{\mu}eV Range with a Coaxial Dish Antenna},''
  \href{http://dx.doi.org/10.1103/PhysRevLett.132.131004}{{\em Phys. Rev.
  Lett.} {\bfseries 132} no.~13, (2024) 131004},
  \href{http://arxiv.org/abs/2310.13891}{{\ttfamily arXiv:2310.13891
  [hep-ex]}}.

\bibitem{CAST:2024eil}
{\bfseries CAST} Collaboration, K.~Altenm\"uller {\em et~al.}, ``{A new upper
  limit on the axion-photon coupling with an extended CAST run with a Xe-based
  Micromegas detector},'' \href{http://arxiv.org/abs/2406.16840}{{\ttfamily
  arXiv:2406.16840 [hep-ex]}}.

\bibitem{Friel:2024shg}
M.~Friel, J.~W. Gjerloev, S.~Kalia, and A.~Zamora, ``{Search for ultralight
  dark matter in the SuperMAG high-fidelity dataset},''
  \href{http://arxiv.org/abs/2408.16045}{{\ttfamily arXiv:2408.16045
  [hep-ph]}}.

\bibitem{Kalia:2024eml}
S.~Kalia, D.~Budker, D.~F.~J. Kimball, W.~Ji, Z.~Liu, A.~O. Sushkov,
  C.~Timberlake, H.~Ulbricht, A.~Vinante, and T.~Wang, ``{Ultralight dark
  matter detection with levitated ferromagnets},''
  \href{http://arxiv.org/abs/2408.15330}{{\ttfamily arXiv:2408.15330
  [hep-ph]}}.

\bibitem{Cadamuro:2011fd}
D.~Cadamuro and J.~Redondo, ``{Cosmological bounds on pseudo Nambu-Goldstone
  bosons},'' \href{http://dx.doi.org/10.1088/1475-7516/2012/02/032}{{\em JCAP}
  {\bfseries 02} (2012) 032}, \href{http://arxiv.org/abs/1110.2895}{{\ttfamily
  arXiv:1110.2895 [hep-ph]}}.

\bibitem{Depta:2020wmr}
P.~F. Depta, M.~Hufnagel, and K.~Schmidt-Hoberg, ``{Robust cosmological
  constraints on axion-like particles},''
  \href{http://dx.doi.org/10.1088/1475-7516/2020/05/009}{{\em JCAP} {\bfseries
  05} (2020) 009}, \href{http://arxiv.org/abs/2002.08370}{{\ttfamily
  arXiv:2002.08370 [hep-ph]}}.

\bibitem{Bauer:2017ris}
M.~Bauer, M.~Neubert, and A.~Thamm, ``{Collider Probes of Axion-Like
  Particles},'' \href{http://dx.doi.org/10.1007/JHEP12(2017)044}{{\em JHEP}
  {\bfseries 12} (2017) 044}, \href{http://arxiv.org/abs/1708.00443}{{\ttfamily
  arXiv:1708.00443 [hep-ph]}}.

\bibitem{Vafa:2005ui}
C.~Vafa, ``{The String landscape and the swampland},''
  \href{http://arxiv.org/abs/hep-th/0509212}{{\ttfamily arXiv:hep-th/0509212}}.

\bibitem{Arkani-Hamed:2006emk}
N.~Arkani-Hamed, L.~Motl, A.~Nicolis, and C.~Vafa, ``{The String landscape,
  black holes and gravity as the weakest force},''
  \href{http://dx.doi.org/10.1088/1126-6708/2007/06/060}{{\em JHEP} {\bfseries
  06} (2007) 060}, \href{http://arxiv.org/abs/hep-th/0601001}{{\ttfamily
  arXiv:hep-th/0601001}}.

\bibitem{Ooguri:2006in}
H.~Ooguri and C.~Vafa, ``{On the Geometry of the String Landscape and the
  Swampland},'' \href{http://dx.doi.org/10.1016/j.nuclphysb.2006.10.033}{{\em
  Nucl. Phys. B} {\bfseries 766} (2007) 21--33},
  \href{http://arxiv.org/abs/hep-th/0605264}{{\ttfamily arXiv:hep-th/0605264}}.

\bibitem{Banks:2010zn}
T.~Banks and N.~Seiberg, ``{Symmetries and Strings in Field Theory and
  Gravity},'' \href{http://dx.doi.org/10.1103/PhysRevD.83.084019}{{\em Phys.
  Rev. D} {\bfseries 83} (2011) 084019},
  \href{http://arxiv.org/abs/1011.5120}{{\ttfamily arXiv:1011.5120 [hep-th]}}.

\bibitem{Ooguri:2016pdq}
H.~Ooguri and C.~Vafa, ``{Non-supersymmetric AdS and the Swampland},''
  \href{http://dx.doi.org/10.4310/ATMP.2017.v21.n7.a8}{{\em Adv. Theor. Math.
  Phys.} {\bfseries 21} (2017) 1787--1801},
  \href{http://arxiv.org/abs/1610.01533}{{\ttfamily arXiv:1610.01533
  [hep-th]}}.

\bibitem{Witten:1984dg}
E.~Witten, ``{Some Properties of O(32) Superstrings},''
  \href{http://dx.doi.org/10.1016/0370-2693(84)90422-2}{{\em Phys. Lett. B}
  {\bfseries 149} (1984) 351--356}.

\bibitem{Kim:1988dd}
J.~E. Kim, ``{The Strong {CP} Problem in Orbifold Compactifications and an
  SU(3) X SU(2) X U(1)-$n$ Model},''
  \href{http://dx.doi.org/10.1016/0370-2693(88)90678-8}{{\em Phys. Lett. B}
  {\bfseries 207} (1988) 434--440}.

\bibitem{Choi:1985bz}
K.~Choi and J.~E. Kim, ``{Compactification and Axions in E(8) x E(8)-prime
  Superstring Models},''
  \href{http://dx.doi.org/10.1016/0370-2693(85)90693-8}{{\em Phys. Lett. B}
  {\bfseries 165} (1985) 71--75}.

\bibitem{Choi:1985je}
K.~Choi and J.~E. Kim, ``{Harmful Axions in Superstring Models},''
  \href{http://dx.doi.org/10.1016/0370-2693(85)90416-2}{{\em Phys. Lett. B}
  {\bfseries 154} (1985) 393}. [Erratum: Phys.Lett.B 156, 452 (1985)].

\bibitem{Derendinger:1985cv}
J.~P. Derendinger, L.~E. Ibanez, and H.~P. Nilles, ``{On the Low-Energy Limit
  of Superstring Theories},''
  \href{http://dx.doi.org/10.1016/0550-3213(86)90396-2}{{\em Nucl. Phys. B}
  {\bfseries 267} (1986) 365--414}.

\bibitem{Banks:1996ss}
T.~Banks and M.~Dine, ``{Couplings and scales in strongly coupled heterotic
  string theory},'' \href{http://dx.doi.org/10.1016/0550-3213(96)00457-9}{{\em
  Nucl. Phys. B} {\bfseries 479} (1996) 173--196},
  \href{http://arxiv.org/abs/hep-th/9605136}{{\ttfamily arXiv:hep-th/9605136}}.

\bibitem{Choi:1997an}
K.~Choi, ``{Axions and the strong CP problem in M theory},''
  \href{http://dx.doi.org/10.1103/PhysRevD.56.6588}{{\em Phys. Rev. D}
  {\bfseries 56} (1997) 6588--6600},
  \href{http://arxiv.org/abs/hep-th/9706171}{{\ttfamily arXiv:hep-th/9706171}}.

\bibitem{Choi:2011xt}
K.~Choi, K.~S. Jeong, K.-I. Okumura, and M.~Yamaguchi, ``{Mixed Mediation of
  Supersymmetry Breaking with Anomalous U(1) Gauge Symmetry},''
  \href{http://dx.doi.org/10.1007/JHEP06(2011)049}{{\em JHEP} {\bfseries 06}
  (2011) 049}, \href{http://arxiv.org/abs/1104.3274}{{\ttfamily arXiv:1104.3274
  [hep-ph]}}.

\bibitem{Buchbinder:2014qca}
E.~I. Buchbinder, A.~Constantin, and A.~Lukas, ``{Heterotic QCD axion},''
  \href{http://dx.doi.org/10.1103/PhysRevD.91.046010}{{\em Phys. Rev. D}
  {\bfseries 91} no.~4, (2015) 046010},
  \href{http://arxiv.org/abs/1412.8696}{{\ttfamily arXiv:1412.8696 [hep-th]}}.

\bibitem{Choi:2014uaa}
K.~Choi, K.~S. Jeong, and M.-S. Seo, ``{String theoretic QCD axions in the
  light of PLANCK and BICEP2},''
  \href{http://dx.doi.org/10.1007/JHEP07(2014)092}{{\em JHEP} {\bfseries 07}
  (2014) 092}, \href{http://arxiv.org/abs/1404.3880}{{\ttfamily arXiv:1404.3880
  [hep-th]}}.

\bibitem{Im:2019cnl}
S.~H. Im, H.~P. Nilles, and M.~Olechowski, ``{Axion clockworks from heterotic
  M-theory: the QCD-axion and its ultra-light companion},''
  \href{http://dx.doi.org/10.1007/JHEP10(2019)159}{{\em JHEP} {\bfseries 10}
  (2019) 159}, \href{http://arxiv.org/abs/1906.11851}{{\ttfamily
  arXiv:1906.11851 [hep-th]}}.

\bibitem{Conlon:2006tq}
J.~P. Conlon, ``{The QCD axion and moduli stabilisation},''
  \href{http://dx.doi.org/10.1088/1126-6708/2006/05/078}{{\em JHEP} {\bfseries
  05} (2006) 078}, \href{http://arxiv.org/abs/hep-th/0602233}{{\ttfamily
  arXiv:hep-th/0602233}}.

\bibitem{Cicoli:2012sz}
M.~Cicoli, M.~Goodsell, and A.~Ringwald, ``{The type IIB string axiverse and
  its low-energy phenomenology},''
  \href{http://dx.doi.org/10.1007/JHEP10(2012)146}{{\em JHEP} {\bfseries 10}
  (2012) 146}, \href{http://arxiv.org/abs/1206.0819}{{\ttfamily arXiv:1206.0819
  [hep-th]}}.

\bibitem{Hebecker:2018yxs}
A.~Hebecker, S.~Leonhardt, J.~Moritz, and A.~Westphal, ``{Thraxions: Ultralight
  Throat Axions},'' \href{http://dx.doi.org/10.1007/JHEP04(2019)158}{{\em JHEP}
  {\bfseries 04} (2019) 158}, \href{http://arxiv.org/abs/1812.03999}{{\ttfamily
  arXiv:1812.03999 [hep-th]}}.

\bibitem{Demirtas:2018akl}
M.~Demirtas, C.~Long, L.~McAllister, and M.~Stillman, ``{The Kreuzer-Skarke
  Axiverse},'' \href{http://dx.doi.org/10.1007/JHEP04(2020)138}{{\em JHEP}
  {\bfseries 04} (2020) 138}, \href{http://arxiv.org/abs/1808.01282}{{\ttfamily
  arXiv:1808.01282 [hep-th]}}.

\bibitem{Halverson:2019cmy}
J.~Halverson, C.~Long, B.~Nelson, and G.~Salinas, ``{Towards string theory
  expectations for photon couplings to axionlike particles},''
  \href{http://dx.doi.org/10.1103/PhysRevD.100.106010}{{\em Phys. Rev. D}
  {\bfseries 100} no.~10, (2019) 106010},
  \href{http://arxiv.org/abs/1909.05257}{{\ttfamily arXiv:1909.05257
  [hep-th]}}.

\bibitem{Mehta:2021pwf}
V.~M. Mehta, M.~Demirtas, C.~Long, D.~J.~E. Marsh, L.~McAllister, and M.~J.
  Stott, ``{Superradiance in string theory},''
  \href{http://dx.doi.org/10.1088/1475-7516/2021/07/033}{{\em JCAP} {\bfseries
  07} (2021) 033}, \href{http://arxiv.org/abs/2103.06812}{{\ttfamily
  arXiv:2103.06812 [hep-th]}}.

\bibitem{Demirtas:2021gsq}
M.~Demirtas, N.~Gendler, C.~Long, L.~McAllister, and J.~Moritz, ``{PQ
  axiverse},'' \href{http://dx.doi.org/10.1007/JHEP06(2023)092}{{\em JHEP}
  {\bfseries 06} (2023) 092}, \href{http://arxiv.org/abs/2112.04503}{{\ttfamily
  arXiv:2112.04503 [hep-th]}}.

\bibitem{Foster:2022ajl}
J.~W. Foster, S.~Kumar, B.~R. Safdi, and Y.~Soreq, ``{Dark Grand Unification in
  the axiverse: decaying axion dark matter and spontaneous baryogenesis},''
  \href{http://dx.doi.org/10.1007/JHEP12(2022)119}{{\em JHEP} {\bfseries 12}
  (2022) 119}, \href{http://arxiv.org/abs/2208.10504}{{\ttfamily
  arXiv:2208.10504 [hep-ph]}}.

\bibitem{Gendler:2023hwg}
N.~Gendler, O.~Janssen, M.~Kleban, J.~La~Madrid, and V.~M. Mehta, ``{Axion
  minima in string theory},'' \href{http://arxiv.org/abs/2309.01831}{{\ttfamily
  arXiv:2309.01831 [hep-th]}}.

\bibitem{Gendler:2023kjt}
N.~Gendler, D.~J.~E. Marsh, L.~McAllister, and J.~Moritz, ``{Glimmers from the
  axiverse},'' \href{http://dx.doi.org/10.1088/1475-7516/2024/09/071}{{\em
  JCAP} {\bfseries 09} (2024) 071},
  \href{http://arxiv.org/abs/2309.13145}{{\ttfamily arXiv:2309.13145
  [hep-th]}}.

\bibitem{Reece:2024wrn}
M.~Reece, ``{Extra-Dimensional Axion Expectations},''
  \href{http://arxiv.org/abs/2406.08543}{{\ttfamily arXiv:2406.08543
  [hep-ph]}}.

\bibitem{Gendler:2024adn}
N.~Gendler and D.~J.~E. Marsh, ``{QCD Axion Dark Matter in String Theory:
  Haloscopes and Helioscopes as Probes of the Landscape},''
  \href{http://arxiv.org/abs/2407.07143}{{\ttfamily arXiv:2407.07143
  [hep-th]}}.

\bibitem{Barr:1981qv}
S.~M. Barr, ``{A New Symmetry Breaking Pattern for SO(10) and Proton Decay},''
  \href{http://dx.doi.org/10.1016/0370-2693(82)90966-2}{{\em Phys. Lett. B}
  {\bfseries 112} (1982) 219--222}.

\bibitem{Derendinger:1983aj}
J.~P. Derendinger, J.~E. Kim, and D.~V. Nanopoulos, ``{Anti-SU(5)},''
  \href{http://dx.doi.org/10.1016/0370-2693(84)91238-3}{{\em Phys. Lett. B}
  {\bfseries 139} (1984) 170--176}.

\bibitem{Pati:1974yy}
J.~C. Pati and A.~Salam, ``{Lepton Number as the Fourth Color},''
  \href{http://dx.doi.org/10.1103/PhysRevD.10.275}{{\em Phys. Rev. D}
  {\bfseries 10} (1974) 275--289}. [Erratum: Phys.Rev.D 11, 703--703 (1975)].

\bibitem{Babu:1985gi}
K.~S. Babu, X.-G. He, and S.~Pakvasa, ``{Neutrino Masses and Proton Decay Modes
  in SU(3) X SU(3) X SU(3) Trinification},''
  \href{http://dx.doi.org/10.1103/PhysRevD.33.763}{{\em Phys. Rev. D}
  {\bfseries 33} (1986) 763}.

\bibitem{Kawamura:1999nj}
Y.~Kawamura, ``{Gauge symmetry breaking from extra space S**1 / Z(2)},''
  \href{http://dx.doi.org/10.1143/PTP.103.613}{{\em Prog. Theor. Phys.}
  {\bfseries 103} (2000) 613--619},
  \href{http://arxiv.org/abs/hep-ph/9902423}{{\ttfamily arXiv:hep-ph/9902423}}.

\bibitem{Hall:2001pg}
L.~J. Hall and Y.~Nomura, ``{Gauge unification in higher dimensions},''
  \href{http://dx.doi.org/10.1103/PhysRevD.64.055003}{{\em Phys. Rev. D}
  {\bfseries 64} (2001) 055003},
  \href{http://arxiv.org/abs/hep-ph/0103125}{{\ttfamily arXiv:hep-ph/0103125}}.

\bibitem{Hall:2001tn}
L.~J. Hall, H.~Murayama, and Y.~Nomura, ``{Wilson lines and symmetry breaking
  on orbifolds},'' \href{http://dx.doi.org/10.1016/S0550-3213(02)00816-7}{{\em
  Nucl. Phys. B} {\bfseries 645} (2002) 85--104},
  \href{http://arxiv.org/abs/hep-th/0107245}{{\ttfamily arXiv:hep-th/0107245}}.

\bibitem{Hall:2001xb}
L.~J. Hall and Y.~Nomura, ``{Gauge coupling unification from unified theories
  in higher dimensions},''
  \href{http://dx.doi.org/10.1103/PhysRevD.65.125012}{{\em Phys. Rev. D}
  {\bfseries 65} (2002) 125012},
  \href{http://arxiv.org/abs/hep-ph/0111068}{{\ttfamily arXiv:hep-ph/0111068}}.

\bibitem{Hebecker:2001jb}
A.~Hebecker and J.~March-Russell, ``{The structure of GUT breaking by
  orbifolding},'' \href{http://dx.doi.org/10.1016/S0550-3213(02)00016-0}{{\em
  Nucl. Phys. B} {\bfseries 625} (2002) 128--150},
  \href{http://arxiv.org/abs/hep-ph/0107039}{{\ttfamily arXiv:hep-ph/0107039}}.

\bibitem{Hebecker:2001wq}
A.~Hebecker and J.~March-Russell, ``{A Minimal S**1 / (Z(2) x Z-prime (2))
  orbifold GUT},'' \href{http://dx.doi.org/10.1016/S0550-3213(01)00374-1}{{\em
  Nucl. Phys. B} {\bfseries 613} (2001) 3--16},
  \href{http://arxiv.org/abs/hep-ph/0106166}{{\ttfamily arXiv:hep-ph/0106166}}.

\bibitem{Altarelli:2001qj}
G.~Altarelli and F.~Feruglio, ``{SU(5) grand unification in extra dimensions
  and proton decay},''
  \href{http://dx.doi.org/10.1016/S0370-2693(01)00650-5}{{\em Phys. Lett. B}
  {\bfseries 511} (2001) 257--264},
  \href{http://arxiv.org/abs/hep-ph/0102301}{{\ttfamily arXiv:hep-ph/0102301}}.

\bibitem{Schellekens:1989qb}
A.~N. Schellekens, ``{Electric Charge Quantization in String Theory},''
  \href{http://dx.doi.org/10.1016/0370-2693(90)91190-M}{{\em Phys. Lett. B}
  {\bfseries 237} (1990) 363--369}.

\bibitem{Narain:1986qm}
K.~S. Narain, M.~H. Sarmadi, and C.~Vafa, ``{Asymmetric Orbifolds},''
  \href{http://dx.doi.org/10.1016/0550-3213(87)90228-8}{{\em Nucl. Phys. B}
  {\bfseries 288} (1987) 551}.

\bibitem{Wen:1985qj}
X.-G. Wen and E.~Witten, ``{Electric and Magnetic Charges in Superstring
  Models},'' \href{http://dx.doi.org/10.1016/0550-3213(85)90592-9}{{\em Nucl.
  Phys. B} {\bfseries 261} (1985) 651--677}.

\bibitem{Font:1990uw}
A.~Font, L.~E. Ibanez, and F.~Quevedo, ``{Higher Level {Kac-Moody} String
  Models and Their Phenomenological Implications},''
  \href{http://dx.doi.org/10.1016/0550-3213(90)90393-R}{{\em Nucl. Phys. B}
  {\bfseries 345} (1990) 389--430}.

\bibitem{Cordova:2018cvg}
C.~C\'ordova, T.~T. Dumitrescu, and K.~Intriligator, ``{Exploring 2-Group
  Global Symmetries},'' \href{http://dx.doi.org/10.1007/JHEP02(2019)184}{{\em
  JHEP} {\bfseries 02} (2019) 184},
  \href{http://arxiv.org/abs/1802.04790}{{\ttfamily arXiv:1802.04790
  [hep-th]}}.

\bibitem{Dine:1987xk}
M.~Dine, N.~Seiberg, and E.~Witten, ``{Fayet-Iliopoulos Terms in String
  Theory},'' \href{http://dx.doi.org/10.1016/0550-3213(87)90395-6}{{\em Nucl.
  Phys. B} {\bfseries 289} (1987) 589--598}.

\bibitem{Wen:1985jz}
X.~G. Wen and E.~Witten, ``{World Sheet Instantons and the {Peccei-Quinn}
  Symmetry},'' \href{http://dx.doi.org/10.1016/0370-2693(86)91587-X}{{\em Phys.
  Lett. B} {\bfseries 166} (1986) 397--401}.

\bibitem{Dienes:1995sq}
K.~R. Dienes, A.~E. Faraggi, and J.~March-Russell, ``{String unification,
  higher level gauge symmetries, and exotic hypercharge normalizations},''
  \href{http://dx.doi.org/10.1016/0550-3213(96)00085-5}{{\em Nucl. Phys. B}
  {\bfseries 467} (1996) 44--99},
  \href{http://arxiv.org/abs/hep-th/9510223}{{\ttfamily arXiv:hep-th/9510223}}.

\bibitem{Farina:2016tgd}
M.~Farina, D.~Pappadopulo, F.~Rompineve, and A.~Tesi, ``{The photo-philic QCD
  axion},'' \href{http://dx.doi.org/10.1007/JHEP01(2017)095}{{\em JHEP}
  {\bfseries 01} (2017) 095}, \href{http://arxiv.org/abs/1611.09855}{{\ttfamily
  arXiv:1611.09855 [hep-ph]}}.

\bibitem{Gavela:2023tzu}
B.~Gavela, P.~Qu\'\i{}lez, and M.~Ramos, ``{The QCD axion sum rule},''
  \href{http://dx.doi.org/10.1007/JHEP04(2024)056}{{\em JHEP} {\bfseries 04}
  (2024) 056}, \href{http://arxiv.org/abs/2305.15465}{{\ttfamily
  arXiv:2305.15465 [hep-ph]}}.

\bibitem{Goodsell:2011wn}
M.~Goodsell, S.~Ramos-Sanchez, and A.~Ringwald, ``{Kinetic Mixing of U(1)s in
  Heterotic Orbifolds},'' \href{http://dx.doi.org/10.1007/JHEP01(2012)021}{{\em
  JHEP} {\bfseries 01} (2012) 021},
  \href{http://arxiv.org/abs/1110.6901}{{\ttfamily arXiv:1110.6901 [hep-th]}}.

\bibitem{Capozzi:2020cbu}
F.~Capozzi and G.~Raffelt, ``{Axion and neutrino bounds improved with new
  calibrations of the tip of the red-giant branch using geometric distance
  determinations},'' \href{http://dx.doi.org/10.1103/PhysRevD.102.083007}{{\em
  Phys. Rev. D} {\bfseries 102} no.~8, (2020) 083007},
  \href{http://arxiv.org/abs/2007.03694}{{\ttfamily arXiv:2007.03694
  [astro-ph.SR]}}.

\bibitem{Antel:2023hkf}
C.~Antel {\em et~al.}, ``{Feebly-interacting particles: FIPs 2022 Workshop
  Report},'' \href{http://dx.doi.org/10.1140/epjc/s10052-023-12168-5}{{\em Eur.
  Phys. J. C} {\bfseries 83} no.~12, (2023) 1122},
  \href{http://arxiv.org/abs/2305.01715}{{\ttfamily arXiv:2305.01715
  [hep-ph]}}.

\bibitem{AxionLimits}
C.~O'Hare, ``cajohare/axionlimits: Axionlimits.''
  \url{https://cajohare.github.io/AxionLimits/}, July, 2020.

\bibitem{Liu:2021zlt}
T.~Liu, X.~Lou, and J.~Ren, ``{Pulsar Polarization Arrays},''
  \href{http://dx.doi.org/10.1103/PhysRevLett.130.121401}{{\em Phys. Rev.
  Lett.} {\bfseries 130} no.~12, (2023) 121401},
  \href{http://arxiv.org/abs/2111.10615}{{\ttfamily arXiv:2111.10615
  [astro-ph.HE]}}.

\bibitem{Gan:2023swl}
X.~Gan, L.-T. Wang, and H.~Xiao, ``{Detecting axion dark matter with black hole
  polarimetry},'' \href{http://dx.doi.org/10.1103/PhysRevD.110.063039}{{\em
  Phys. Rev. D} {\bfseries 110} no.~6, (2024) 063039},
  \href{http://arxiv.org/abs/2311.02149}{{\ttfamily arXiv:2311.02149
  [hep-ph]}}.

\bibitem{Carroll:1989vb}
S.~M. Carroll, G.~B. Field, and R.~Jackiw, ``{Limits on a Lorentz and Parity
  Violating Modification of Electrodynamics},''
  \href{http://dx.doi.org/10.1103/PhysRevD.41.1231}{{\em Phys. Rev. D}
  {\bfseries 41} (1990) 1231}.

\bibitem{Harari:1992ea}
D.~Harari and P.~Sikivie, ``{Effects of a Nambu-Goldstone boson on the
  polarization of radio galaxies and the cosmic microwave background},''
  \href{http://dx.doi.org/10.1016/0370-2693(92)91363-E}{{\em Phys. Lett. B}
  {\bfseries 289} (1992) 67--72}.

\bibitem{Lue:1998mq}
A.~Lue, L.-M. Wang, and M.~Kamionkowski, ``{Cosmological signature of new
  parity violating interactions},''
  \href{http://dx.doi.org/10.1103/PhysRevLett.83.1506}{{\em Phys. Rev. Lett.}
  {\bfseries 83} (1999) 1506--1509},
  \href{http://arxiv.org/abs/astro-ph/9812088}{{\ttfamily
  arXiv:astro-ph/9812088}}.

\bibitem{Strominger:1990et}
A.~Strominger, ``{Heterotic solitons},''
  \href{http://dx.doi.org/10.1016/0550-3213(90)90599-9}{{\em Nucl. Phys. B}
  {\bfseries 343} (1990) 167--184}. [Erratum: Nucl.Phys.B 353, 565--565
  (1991)].

\end{thebibliography}\endgroup

\end{document}